\documentclass[11pt, twoside]{predoc2}
\usepackage[utf8]{inputenc}
\usepackage{distoperators}
\usepackage{newtxtext}
\usepackage[scaled=.95]{cabin}
\usepackage[varqu,varl]{inconsolata} 
\usepackage[title]{appendix}

\usepackage{enumitem}
\usepackage[T1]{fontenc}

\usepackage{todonotes}
\RequirePackage{natbib}

\usepackage{xcolor}
\definecolor{darkblue}{RGB}{0, 0, 128}

\usepackage{afterpage}
\usepackage{placeins}
\usepackage{subcaption}

\makeatletter
\newcommand{\subalign}[1]{%
  \vcenter{%
    \Let@ \restore@math@cr \default@tag
    \baselineskip\fontdimen10 \scriptfont\tw@
    \advance\baselineskip\fontdimen12 \scriptfont\tw@
    \lineskip\thr@@\fontdimen8 \scriptfont\thr@@
    \lineskiplimit\lineskip
    \ialign{\hfil$\m@th\scriptstyle##$&$\m@th\scriptstyle{}##$\hfil\crcr
      #1\crcr
    }%
  }%
}
\makeatother

\newcommand{\given}{\,|\,}
\newcommand{\diff}{\mathrm{d}}

\newcommand{\yobs}{y_{\text{obs}}}
\newcommand{\Nobs}{N_{\text{obs}}}

\newcommand{\new}[1]{#1}

\usepackage{url}
\title{Posterior SBC: Simulation-Based Calibration Checking Conditional on Data}
\author[1,*]{Teemu Säilynoja}
\author[2]{Marvin Schmitt}
\author[3]{Paul-Christian Bürkner}
\author[1]{Aki Vehtari}
\affil[1]{Department of Computer Science, Aalto University, Finland}

\affil[2]{Cluster of Excellence SimTech, University of Stuttgart, Germany}

\affil[3]{Department of statistics, University of Dortmund, Germany}

\affil[*]{Corresponding author: teemu.sailynoja@aalto.fi}
\date{\today}

\makeatletter
\def\runauthor{T. Säilynoja et al.}
\def\runtitle{Posterior Simulation-Based Calibration Checking}
\makeatother

\begin{document}

\maketitle

\begin{abstract}
Simulation-based calibration checking (SBC) refers to the validation of an inference algorithm and model implementation through repeated inference on data simulated from a generative model. In the original and commonly used approach, the generative model uses parameters drawn from the prior, and thus the approach is testing whether the inference works for simulated data generated with parameter values plausible under that prior. This approach is natural and desirable when we want to test whether the inference works for a wide range of datasets we might observe. However, after observing data, we are interested in answering whether the inference works conditional on that particular data. In this paper, we propose posterior SBC and demonstrate how it can be used to validate the inference conditionally on observed data. We illustrate the utility of posterior SBC in three case studies: (1) A simple multilevel model; (2) a model that is governed by differential equations; and (3) a joint integrative neuroscience model which is approximated via amortized Bayesian inference with neural networks.

\end{abstract}

\section{Introduction}

\begin{figure}[t]
    \centering
    \begin{subfigure}{.45\linewidth}\includegraphics[width=\linewidth]{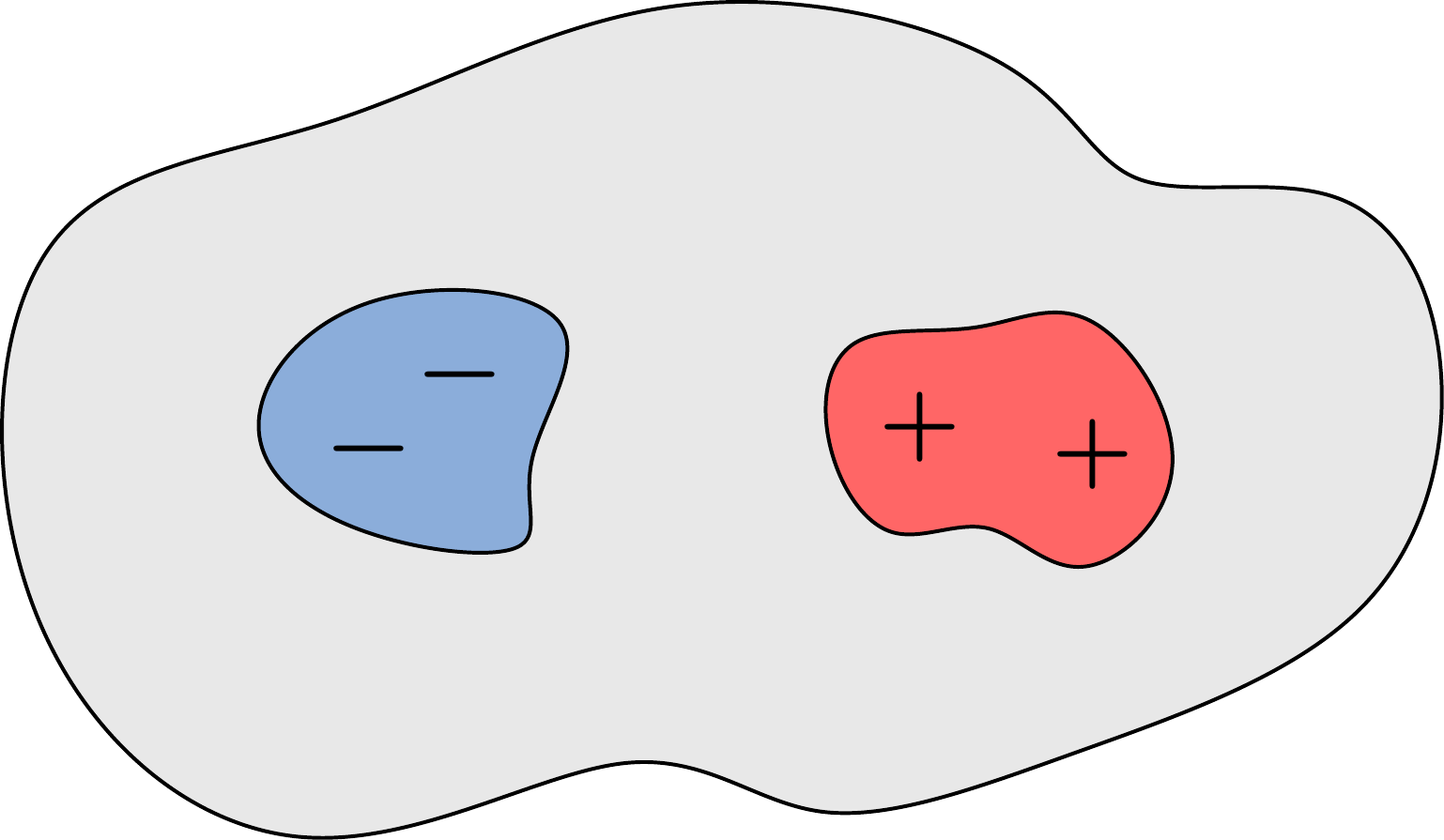}
    \caption{Two cancelling biases}
    \end{subfigure}
    \hspace*{0.05\linewidth}
    \begin{subfigure}{.45\linewidth}
        \includegraphics[width=\linewidth]{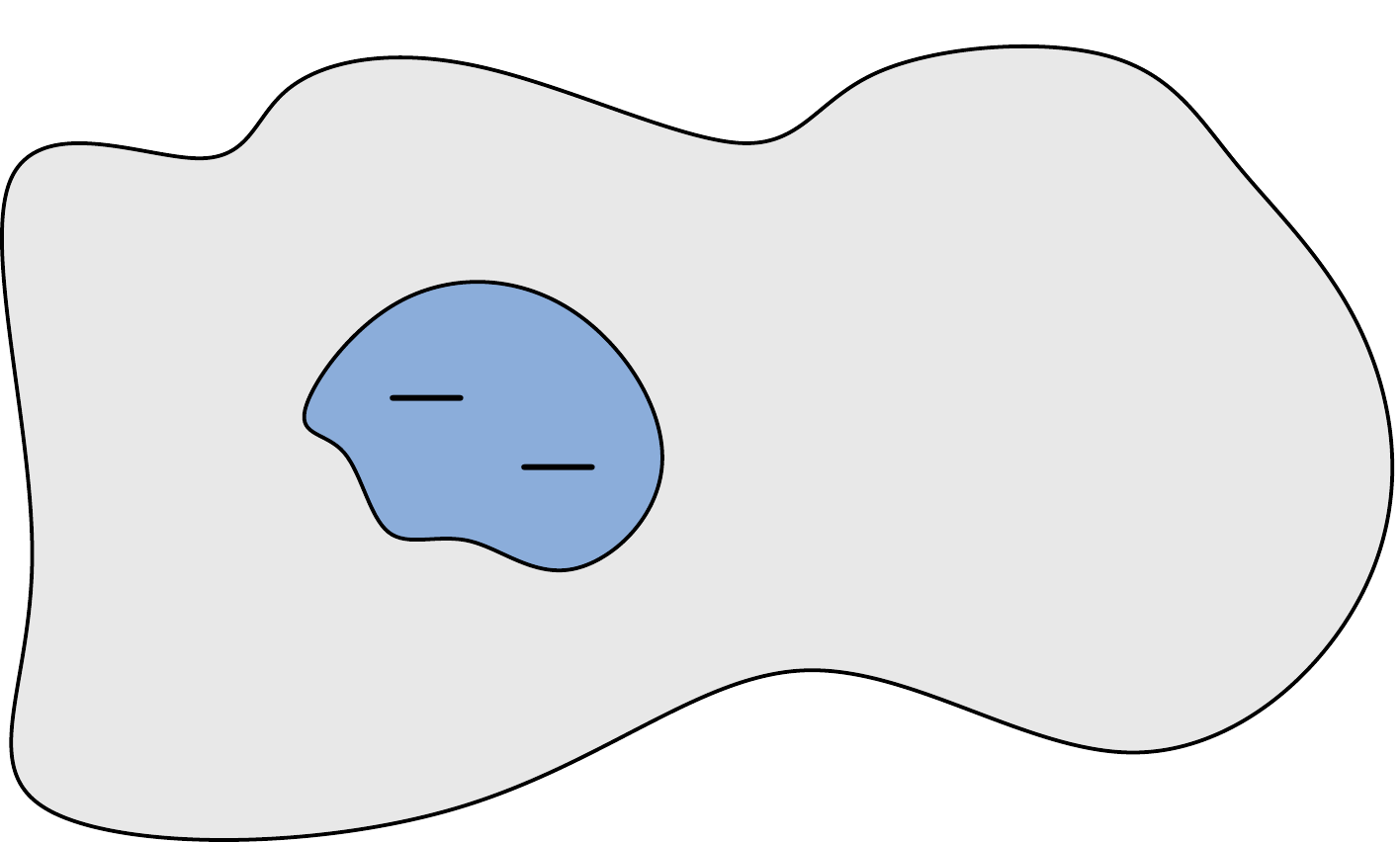}
    \caption{One problematic region}
    \end{subfigure}
    \caption{Two conceptual illustrations of a model parameter space, with gray area denoting the prior. In (a), the colored regions are potential posteriors with bias in opposite directions, while the inference is well calibrated for parameter values outside these regions. \new{Prior SBC will not show calibration issues due to cancellation of biases, while posterior SBC would indicate issues for posteriors intersecting the colored regions. In (b), only one region is problematic. Now prior SBC would indicate calibration issues, while posterior SBC would show that the inference is calibrated outside the colored region. In both cases posterior SBC has the benefit of focusing on the region, which matters given the observed dataset.}}
    \label{fig:intro-here-be-dragons}
\end{figure}

To ensure trust on the results of the inference, different forms of calibration checking play important roles in the Bayesian workflow for model building \new{\citep{Gelman_bayesian_2020}}. Simulation-based calibration checking \new{\citep[SBC; ][]{cook_validation_2006,talts_validating_2020,modrak_simulation-based_2023}} is one of these methods and validates the chosen posterior inference algorithm as well as checks for inconsistencies between the model implementation and a possibly separate implementation of the data generating process.
\new{Probabilistic programming software \citep{Strumbelj+etal:2024:software} provide probabilistic programming languages to make it easier to implement probabilistic models, and inference engines that implement various inference algorithms. As the probabilistic programming software and algorithms for probabilistic modelling advance, the threshold for specifying increasingly complicated models gets lowered. However, as the complexity of the models increases, so does the probability of human error when writing the code, which implements the model.
In addition to such model code implementation mistakes, the inference algorithms used for obtaining posterior approximations---even when correctly coded---can work well for one posterior, but suffer from computational difficulties for another. By simply changing the observed data, one may obtain vastly different posterior geometries, favoring one implementation or algorithm over another.}

\new{SBC aims to check that the inference is calibrated, but miscalibration can be caused either a mistake in the model implementation code, a mistake in inference algorithm implementation code, or significant bias in the inference algorithm. The algorithm is considered to be calibrated, if the probability integral transformation (PIT) of the parameters is uniformly distributed, when transformed with respect to the posterior conditional on data generated using those parameters. We review SBC and the definition of calibration in more detail in Section \ref{sec:prior_SBC}.}

\new{In this paper we assume that the both the code for implementing the model, and the inference algorithm implementation are correct and focus on examples were the inference algorithm may produce significantly biased results with finite computation time. Even when the inference algorithm is implemented correctly, it may produce significantly biased inference for some posteriors. For example, many Markov chain Monte Carlo (MCMC) and variational inference approaches have challenges with funnel shaped posteriors \citep[see, e.g.][]{Neal:2003:slice,Yao:2018:did_it_work}, which are common in case of hierarchical models.
The original SBC is limited to checking for calibration on the whole joint distribution of parameters and observations under the specified priors, and thus we call it \emph{prior SBC}. Specifying prior distributions that accurately present the available prior information is often difficult.  
In some cases, the prior may have significant probability mass in parts of the parameter space that generate such data for which the inference algorithm produce significantly biased inference (in finite time). For example, the data generated from a hierarchical model with certain type of parameter values can lead to a funnel shaped posterior which has a challenging geometry for many inference algorithms to explore. Figure \ref{fig:intro-here-be-dragons} illustrates two cases: a) in which the prior space includes two subspaces with canceling biases and SBC indicates that the calibration over the prior is good, and b) in which the prior space includes one subspace with biased inference causing SBC to indicate that the calibration over the prior is not good. However, after observing some data we are not interested in calibration over the prior, but calibration within the region of the posterior.}

\new{In this paper, we introduce \emph{posterior SBC}, a variant of SBC using the self-consistency of Bayesian posterior distribution to validate the model implementation and inference algorithm when conditioned on some fixed set of data. Figure \ref{fig:intro-here-be-dragons} illustrates two cases where, if the posterior is much more concentrated than the prior, posterior SBC checks the calibration of the inference for those regions of the parameter space where it matters and does not care how well the inference algorithm works elsewhere. Prior SBC is useful for algorithm and software developers, while posterior SBC is useful for modellers. As the number of modellers is several orders of magnitude bigger than the number of developers, we expect the posterior SBC to become more popular than prior SBC.}


We next review the method of simulation-based calibration as it has traditionally been operationalized.
In Section~\ref{sec: Posterior SBC}, we introduce posterior SBC. In Section~\ref{Sec: Case Studies} we illustrate the differences between prior and posterior SBC through three case studies including a simple hierarchical model, Lotka-Volterra model, and a drift-diffusion model. \new{In the first two examples we use MCMC and in the last example we use amortized Bayesian inference for which there are no similar convergence diagnostics as for MCMC.} Finally, in Section~\ref{sec: conclusions} we summarize the contributions of the paper and provide an outlook for future research.

\section{Simulation-based calibration checking}
\label{sec:prior_SBC}

\new{Let us consider a Bayesian model of the joint distribution of the data $y$ and parameters $\theta$,
$$
\pi(y,\theta) = \pi(y \given \theta)\pi(\theta),
$$
where $\pi(\theta)$ is the prior distribution of the parameters and $\pi(y\given\theta)$ the likelihood of the data.}

\new{\citet{cook_validation_2006} propose a simulation-based calibration checking method for validating Bayesian inference software designed to fit the model. The key to this method is the following self-consistency property of Bayesian posterior distributions. Let $\theta' \sim \pi(\theta)$, and $y' \sim \pi(y\given\theta')$. Now the pair $(y',\theta')$ represents a draw from the joint distribution $\pi(y,\theta)$, and therefore $\theta'$ is a draw from the posterior distribution, $\pi(\theta\given y')$. This self-consistency property, linking the prior, prior predictive, and posterior distributions, can be summarized with the following \emph{SBC equality}:
\begin{equation}\label{Eq: Prior SBC 1}
    \pi(y',\theta',\theta'') = \pi(\theta''\given y')\pi(y'\given \theta')\pi(\theta') = \pi(\theta''\given y')\pi(\theta'\given y')\pi(y'),
\end{equation}
where $\theta'' \sim \pi(\theta\given y')$ is a posterior draw conditioned on $y'$.}

\new{Given an observation $y'$ and the model implementation, the inference algorithm should be able to sample from the posterior, $\theta''\sim \pi(\theta\given y')$.
Therefore, to verify that the inference algorithm is able to accurately sample from the posterior, we should assess whether the posterior draw $\theta''$ and the prior draw $\theta'$, conditional on $y'\sim\pi(y\given\theta')$, are from the same distribution.}

\begin{figure}
    \centering
    \includegraphics[width=.6\linewidth]{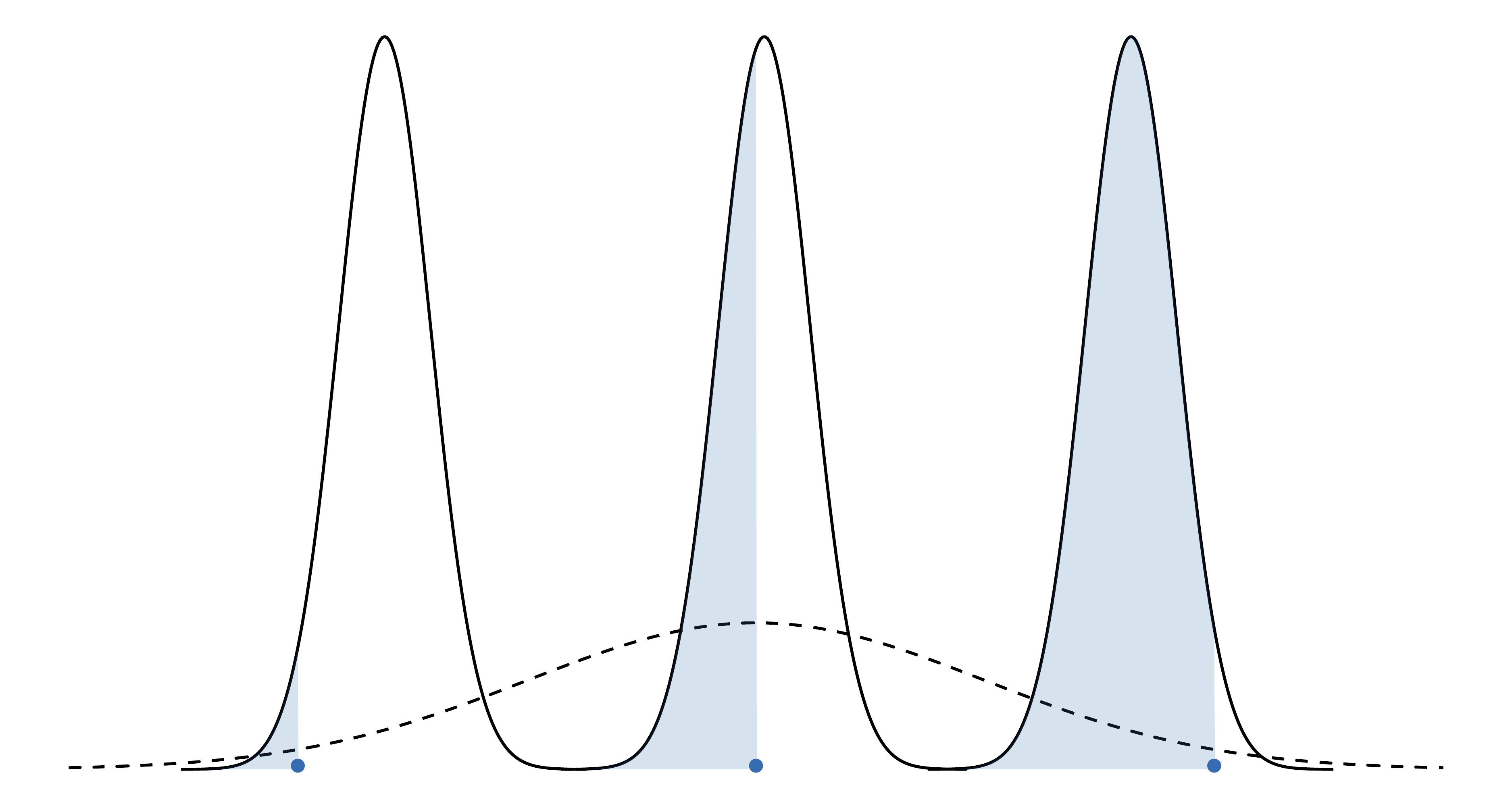}
    \caption{\new{An illustration of how PIT values are computed in prior SBC. The model is a normal distribution $\mathcal N(\theta,\sigma)$ with a known standard deviation $\sigma$. The dashed line shows the prior distribution $p(\theta)$, and solid lines show three posteriors $p(\theta|y_i)$ conditioning on the predictive draws $y_i \sim \pi(y_i\given\theta' _i)$. The PIT value of each prior draw $\theta'_i$, shown as blue dots, equals the respective shaded blue area ($0.03$, $0.43$, and $0.97$ respectively). In practice the PIT values (\ref{eq:pit}) are computed using Monte Carlo draws from the posteriors $\theta''_{i,1},\ldots,\theta''_{i,S} \sim\pi(\theta''\given y_i)$.}}
    \label{fig:intro-sbc_pit}
\end{figure}

To assess whether $\theta'$ and $\theta''$ are from the same distribution conditional on $y$,
\cite{cook_validation_2006} repeatedly draw parameter vectors $\theta'_i$ from the prior $\pi(\theta)$, generate data from the observation model $y_i \sim \pi(y_i\given\theta' _i)$, and then use the algorithm to be validated to sample $\theta''_{i,1},\ldots,\theta''_{i,S} \sim\pi(\theta''\given y_i)$. The authors then propose to compute empirical probability integral transform (PIT) values,
\begin{align}\label{eq:pit}
  u_i = p(\theta_i'' < \theta_i' \given y_i),
\end{align}
which prove to be uniformly distributed as $S\to\infty$. The process is repeated for $i=1,\ldots,N$ and the empirical PIT values are used for testing. Figure~\ref{fig:intro-sbc_pit} shows three iterations of prior draws and their respective PIT relative to the corresponding posterior distributions. \cite{cook_validation_2006} further propose the use of the $\chi^2$-test for the inverse of the normal cumulative distribution function (CDF) of the empirical PIT values. \new{The inference is said to be calibrated if the PIT values pass a uniformity test.}

\cite{cook_validation_2006} did not take into account that the estimated PIT values are discrete when computed with finite sample size $S$, and how the possible correlation in Markov chains affects the results \citep{gelman_correction_2017,talts_validating_2020}.  \citet{talts_validating_2020} propose testing for the discrete uniformity of the empirical PIT values \citep[see also][for improved theory and proofs]{modrak_simulation-based_2023}.
By thinning $\theta_{i, 1}'',\ldots,\theta_{i, S}''$ to remove possible autocorrelation, the uniformity of these discrete empirical PIT values can be tested with the approach presented in \citet{sailynoja_graphical_2022}.

\new{Instead of only assessing the uniformity of the ranks of individual parameters, \citet[][Sections 3.3--3.4]{modrak_simulation-based_2023} propose the use of the joint log-likelihood $\log p(y \given \theta)$ as a test statistic for assessing the calibration. That is, $p(y \given \theta')$ and $p(y \given \theta'')$ are used instead of $\theta'$ and $\theta''$ when computing PIT values with (\ref{eq:pit}). This has the benefit of being a joint function of both data and all parameters. Using the data in the test quantity allows detecting calibration issues rising from the model partially ignoring the data. \citet[][Sections 3.4 and 4.3]{modrak_simulation-based_2023} demonstrate how test quantities based on only parameters can not detect if the model completely ignores all data (the posterior is the prior) or part of the data (e.g. the first data point), but using the log-likelihood as a test quantity indicates the problem. Another advantage of using the log-likelihood as a test quantity is in checking calibration of models with high-dimensional parameter spaces, where modelers might not be particularly interested in gauging the calibration of all the individual parameters, and using the log-likelihood as a test quantity is a natural choice for a joint function of model parameters.
Using the joint log-likelihood also avoids the need of correcting for multiple comparisons from simultaneously assessing the calibration of many individual parameters.}

\new{Performing SBC on the full prior parameter space has three fundamental drawbacks.
First, as shown in the case study of Section~4.1, it is possible that only some parameter values generate such data that lead to a posterior from which the inference method is not able to sample faithfully. If such parameter values occur in a small region of the parameter space, and we average over a much wider prior, the effect in the calibration checking can be minor and prior SBC may miss the miscalibration completely. If such parameter values have high probability mass under the posterior, then posterior SBC shows the miscalibration.}

Second, the time required for running the inference on each SBC iteration is often non-trivial and one needs to consider the tradeoff between their computational budget and the power of the calibration checking.
Third, as visualized in Figure~1, even with unlimited computational resources the result of prior SBC does not guarantee calibration within a specific subregion of the inspected parameter space. There can always be another region with an equally strong but reverse effect on the overall distribution of the PIT values.
As proposed by \cite{modrak_simulation-based_2023}, one solution to this last problem is to employ additional test quantities using functions of parameters and non-monotonic transformations (such as folded rank statistics), that might detect these problematic regions, but coming up with these test quantities can be  difficult.

\section{Posterior simulation-based calibration checking -- conditioned on observed data}\label{sec: Posterior SBC}

Weakly informative priors are commonly used in Bayesian inference due to their desirable properties in real-life applications \citep{gelman_prior_2017}.
However, this class of priors can cause serious issues for interpreting the results of prior SBC. Even though a large portion of the prior mass may lie on a region of the parameter space where the inference works well, \new{smaller regions may exist where the inference method can fail to faithfully present the posterior.} On the one hand, if large enough, these issues may be detected through prior SBC. The modeller could then try to modify the priors or the model to reduce the prior probability of the problematic parameter values. On the other hand, if one already has access to the data that the model will later be used to run inference on, the assessment of the inference algorithm could be focused on the region of the parameter space around the posterior. For inference on the given data, this is arguably the most important region to ensure proper calibration for.

Using the sequential Bayesian updating rule, one can view the posterior as the new prior and perform the entire SBC procedure conditional on the observed data $y_{obs}$.
In other words, by expanding the joint distribution in (\ref{Eq: Prior SBC 1}) to include both the observed data and the predictive draw,
\begin{align}\label{Eq: Posterior SBC}
\pi(\yobs, y,\theta',\theta'') =  \pi(\theta'\given\yobs)\pi(y\given \theta', \yobs)\pi(\theta''\given y, \yobs),
\end{align}
we see that a sample from the old posterior, $\theta'_i \sim \pi(\theta\given y_{obs})$, and $\theta'' \sim \pi(\theta\given y_i,y_{obs})$ drawn from an augmented posterior, need to have the same distribution given the prediction $y_i\sim \pi(y \given \theta'_i)$.
Thus, we operationalize the approach for data conditioned SBC by drawing $\theta'_i \sim \pi(\theta\given\yobs)$, generating posterior predictions $y_i\sim \pi(y \given \theta'_i)$, and then using the inference algorithm to be validated to draw from the new posterior $\theta''_1,\dots \theta''_S \sim \pi(\theta\given y_i,\yobs)$.

\new{The calibration of inference can be checked} by testing the uniformity of the PIT values, $u_i = p(\theta''_i < \theta'_i\given y_i,y_{obs})$, this time computed for the original posterior draws, $\theta'_i$, with regard to the (thinned) augmented posterior draws, $\theta''_{i,1},\dots,\theta''_{i, N}$. We call this variant of SBC using posterior draws \emph{posterior SBC}.

\new{When using prior SBC for validating the inference algorithm, drawing parameter values generatively from the prior is usually assumed to be trivial. In posterior SBC, we need to sample from two different posteriors,  which would usually be achieved using the same algorithm that is being evaluated. Posterior SBC does not require that we are able to get draws from the true posterior. If the sampling fails for either the original posterior or for the augmented posteriors, the SBC equality in Equation \ref{Eq: Posterior SBC}
does not hold, leading to non-uniform PIT values in posterior SBC. While the Bayesian updating is consistent when more data is observed, it is very unlikely that we would encounter such biased inference that would be consistent when conditioned on more data.}

\new{The posterior SBC looks similar to some other methods using draws from the posterior predictive distribution. Posterior predictive checking \citep[PPC; ][]{Box:1980,Rubin:1984,gelman1996posterior,BDA3,gabry_visualization_2019} compares draws from the posterior predictive distribution to the original data. PPC uses the data twice which can lead to optimistic checking unless the test statistic is completely ancillary which is unlikely especially for more flexible models. Parameter recovery experiments generate data from the model using some feasible parameter values, and the posterior is compared to those parameter values visually without a formal test. The crucial part of posterior SBC is to use the chain rule and compare the draws from the original posterior to the augmented posterior which is conditioned on both the original data and the posterior predictive draws.}

\new{A reader may wonder why do we need posterior SBC, when sampling from the original posterior with Markov chain Monte Carlo can be diagnosed with convergence diagnostics such as $\widehat{R}$ \citep{Vehtari+etal:2021:Rhat} and divergences \citep{Betancourt:2017:conceptual_hmc}. \citet[][Appendix C]{Vehtari+etal:2021:Rhat} demonstrate how $\widehat{R}$ fails to diagnose bias in the case of a funnel shaped posterior and very long chains. While in that example Hamiltonian Monte Carlo specific divergence diagnostic did work, it has also been observed to miss convergence issues. Furthermore, additional diagnostics similar to $\widehat{R}$ and the divergence diagnostic are not available for all MCMC methods.
What is more, while convergence diagnostics can indicate issues, they do not indicate the direction or magnitude of the induced bias. For example, it is known that there are also false positive divergence warnings, and in such cases posterior SBC can be used to measure the amount of potential bias.
While the posterior SBC is operationalized by sampling from the posterior, the inference algorithms tested do not need to be (Markov chain) Monte Carlo methods. Any inference algorithm can be tested as long as we can also get draws from the approximated posterior. \citep{Yao:2018:did_it_work} demonstrate the challenges of diagnosing the reliability of variational inference in high dimensions and use prior SBC to check variational inference by drawing from the variational approximation. \citet{Schad+etal:2023:workflow_bayes_factors} discuss problems in estimating Bayes factors and use prior SBC for checking computation of Bayes factors with bridge sampling and Savage-Dickey method. Considering \citep{Yao:2018:did_it_work} and \citet{Schad+etal:2023:workflow_bayes_factors} did consider reliability of computation given the observed data, it would have been sensible to use posterior SBC, instead. In Section \ref{sec:case-study-abi} we illustrate the use posterior SBC for amortized Bayesian inference, for which before this there were no practical inference checking tools.}

\section{Case studies}\label{Sec: Case Studies}

In this section, we present three case-studies demonstrating the differences of prior and posterior SBC. We demonstrate how posterior SBC is better suited for calibration assessment when the modeller is particularly interested in the trustworthiness of the inference with a given dataset. In this case, Prior SBC may be too computationally ineffective to detect calibration issues, or alert of issues that are not present after conditioning on the data.

In Section~\ref{sec:case-study-hm}, we show an example where both of two alternative model implementations exhibit miscalibration. However, the problematic region is relatively small. Prior SBC does not imply calibration issues even with a large number of iterations, while posterior SBC is able to detect the issue.

In Section~\ref{sec:case-study-lv}, we show an example where prior SBC indicates calibration issues with data plausible under prior beliefs of the parameter values. After conditioning the model on an observed dataset, these problems are no longer present. Additionally, in this example, the challenges caused by the chosen diffuse priors make the inference computationally heavy, and thus would make it impractical to iteratively improve the model.

Lastly, in Section~\ref{sec:case-study-abi}, we illustrate the potential of posterior SBC in amortized Bayesian workflows. There, it can be employed as the default data-conditional diagnostic for assessing the trustworthiness of the learned neural posterior approximator with negligible computational overhead.

\subsection{Simple hierarchical model -- choice of model parameterization}\label{sec:case-study-hm}
This case study focuses on a situation where the best choice of model implementation to obtain trustworthy inference depends on the observed data. Hierarchical models often exhibit funnel shaped posterior geometries that are difficult to sample \citep{papaspiliopoulos_general_2007}. Depending on the observed data, the likelihood can contribute to the posterior in vastly different ways, making it near-impossible to choose a model parameterization before observing data.  
\paragraph{Setup}
We study a simple hierarchical model for a dataset with 50 groups each producing five observations,
\begin{align*}
    y_{i,j} &\sim \mathcal N(\mu_j,\sigma^2),\\
    \mu_j &\sim \mathcal N (\mu_0, \tau^2),\\
    \mu_0 &\sim \mathcal N(0,1),\\
    \sigma, \tau &\sim \mathcal N^+(0,1),
\end{align*}
where $j \in \{1,\dots,50\}$, and $i \in \{1,\dots, 5\}$.

The standard normal prior for the population level mean $\mu_0$, as well as the truncated standard normal priors for both the population level variance, $\tau$, and the within group variance, $\sigma$, quantify relatively large prior uncertainty on both the similarity between the groups, and the similarity of the observations within any given group.

We first summarize the results of using prior SBC to investigate the calibration of the posterior inference with two alternative parametrizations of the joint likelihood: the centered parametrization shown above, and the mathematically equivalent non-centered parametrization:
\begin{align*}
    y_{i,j} &\sim \mathcal N(\mu_j,\sigma^2),\\
    \mu_j &= \mu_0 + \tau z_j,\\
    z_j &\sim \mathcal N (0, 1),\\
    \mu_0 &\sim \mathcal N(0,1),\\
    \sigma, \tau &\sim \mathcal N^+(0,1).
\end{align*}
The non-centered parameterization exhibits a more convenient posterior geometry in case of a weakly informative likelihood.

For posterior SBC, we assess the calibration of posterior inference conditional on two datasets, the first generated with $\tau = 0.06$ and $\sigma = 1.96$, and the second with $\tau = 1.96$ and $\sigma = 0.06$. These correspond to the 5th and 95th percentile of the truncated standard normal priors. The first case has a strong prior and a weak likelihood leading to a funnel shaped posterior with centered parameterization. The second case has a weak prior and a strong likelihood leading to a funnel shaped posterior with non-centered parameterization. The population mean was fixed at $\mu_0 = 0$ for both observations.

We implement this model with centered and non-centered parameterizations using the Stan probabilistic programming language \citep{stan_development_team_stan_2023} and run the posterior inference with its default MCMC sampling algorithm, the no-U-turn sampler \citep{hoffman_no-u-turn_2014,stan_development_team_stan_2023}. Despite the slower sampling, we raise the target acceptance ratio, called adapt delta, from $0.80$ to $0.99$, to more reliably sample posteriors with highly varying curvature. In each iteration of both prior and posterior SBC, we sample four MCMC chains of 1000 warm-up draws and 1000 posterior draws. \new{To automate SBC iterations, we used the \texttt{sbc} R package \citep{sbc_package:2024}.}

\paragraph{Results}
\begin{figure}[t]
    \centering
    \begin{subfigure}{.45\linewidth}
        \includegraphics[width=\linewidth]{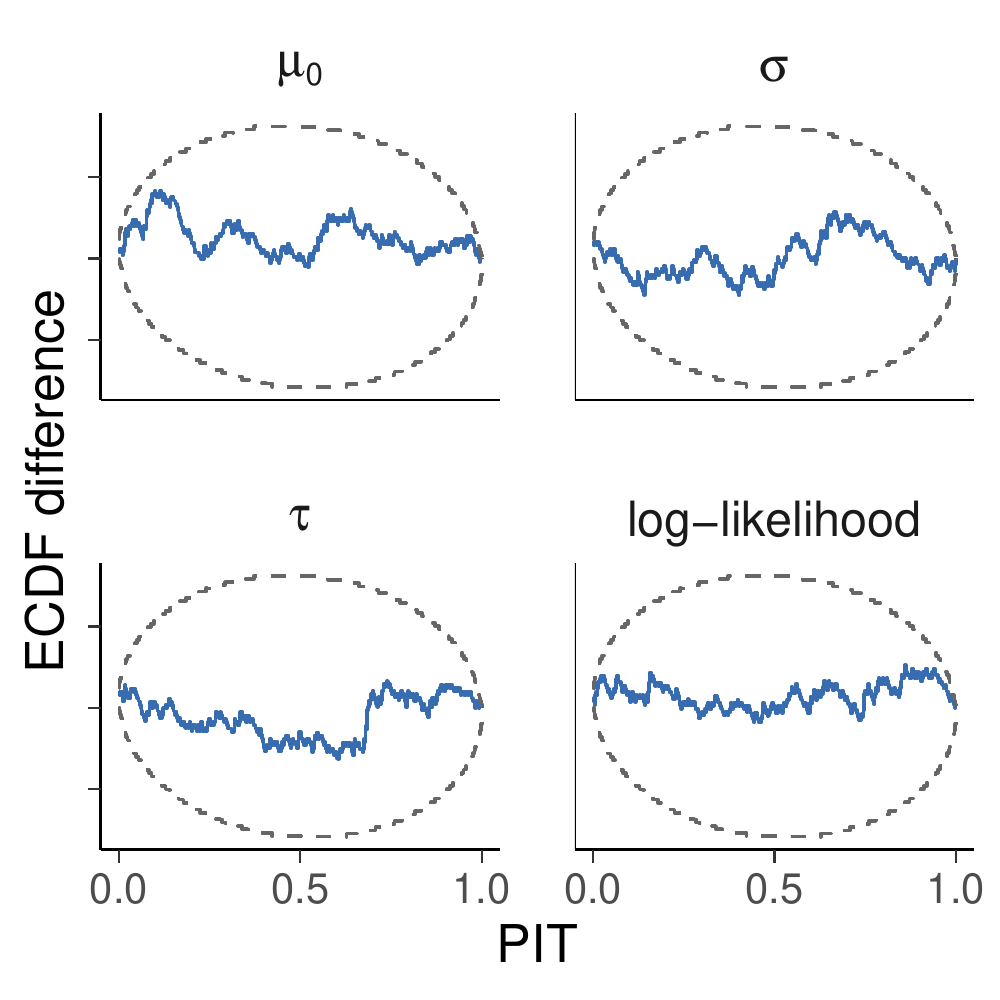}
        \caption{Centered Parameterization}
    \end{subfigure}
    \hspace*{0.05\linewidth}
    \begin{subfigure}{.45\linewidth}
        \includegraphics[width=\linewidth]{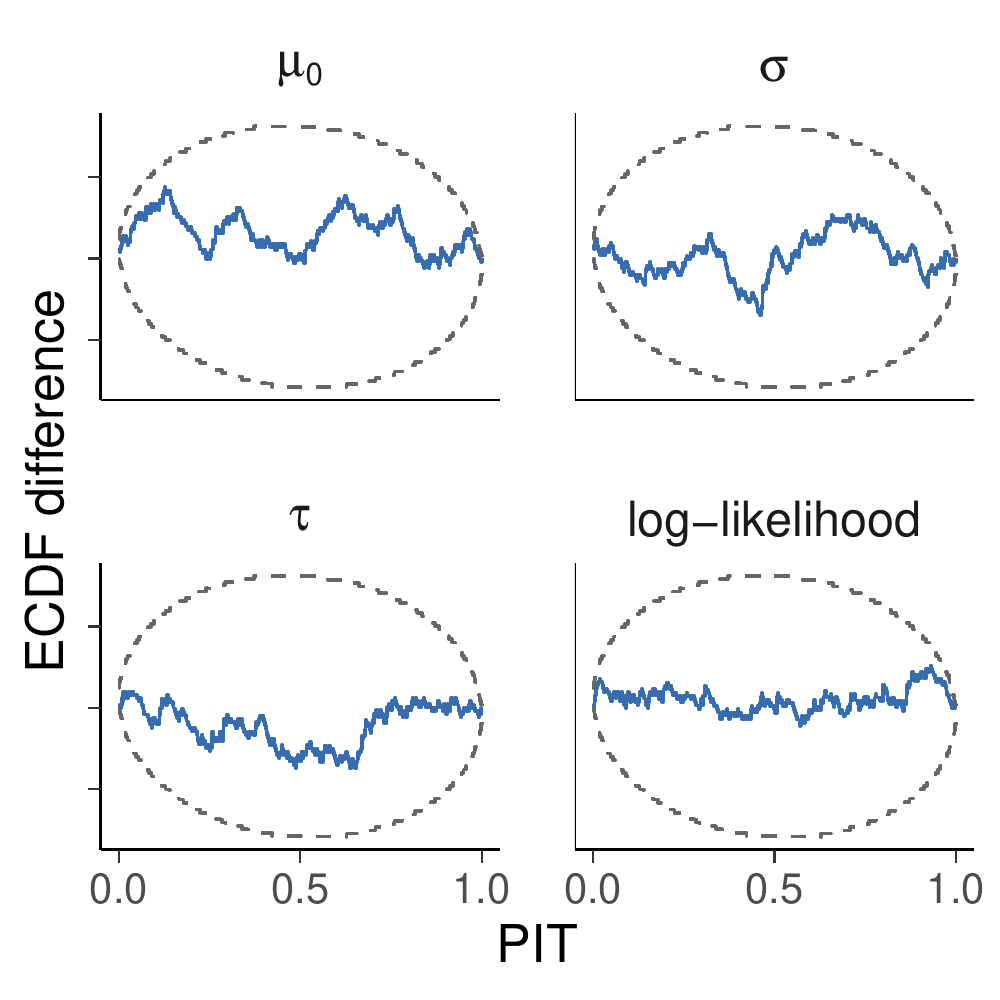}
        \caption{Non-centered Parameterization}
    \end{subfigure}
    \caption{\textit{Prior SBC} checking  for the hierarchical model with (a) the centered and (b) the non-centered parameterization using four different test quantities. \new{Subplots show PIT-ECDF difference plots using four different test quantities and 95\% simultaneous confidence intervals under the assumption of uniformity. As all blue PIT-ECDF difference lines are inside the 95\% simultaneous confidence intervals, the inference seems to be calibrated when SBC iterations are averaged over prior draws.}}
    \label{fig:hm-prior-SBC}
\end{figure}
\begin{figure}
    \centering
    \begin{subfigure}{0.45\linewidth}
        \includegraphics[width=\linewidth]{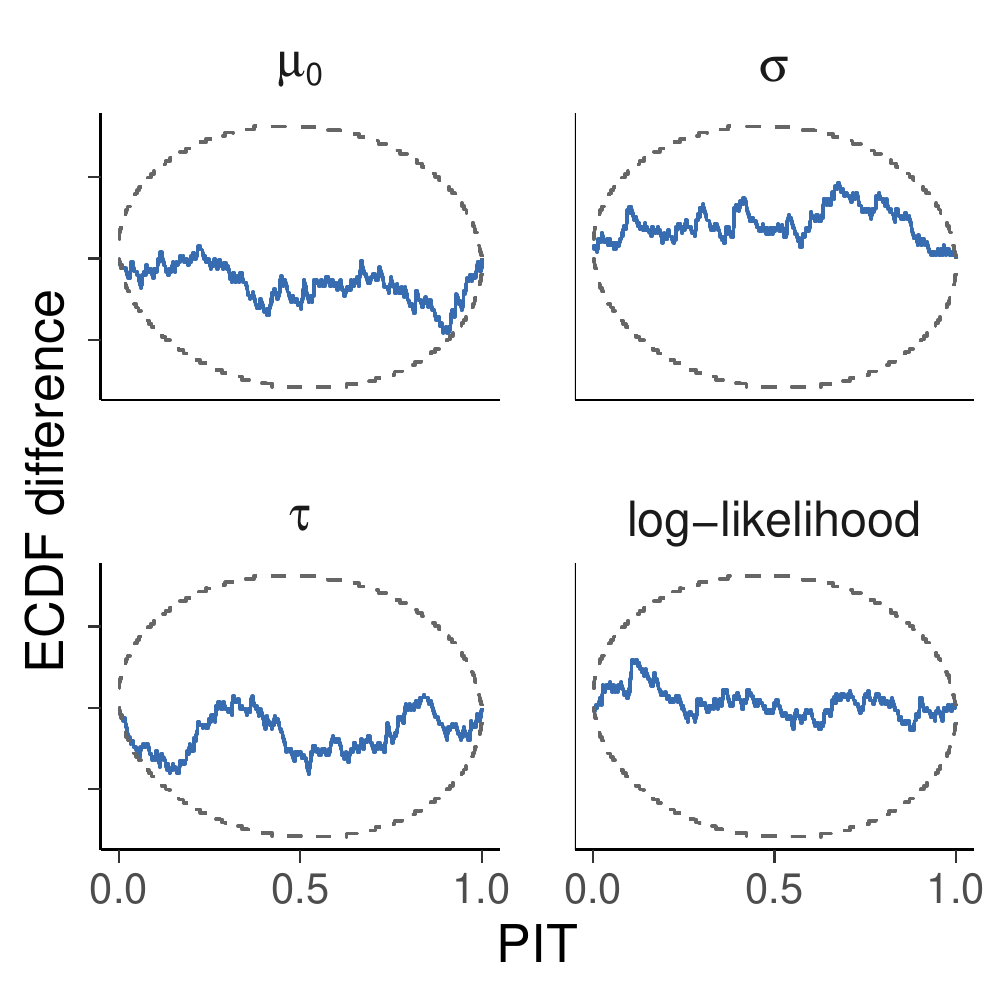}
        \caption{Centered Parameterization}
    \end{subfigure}
    \hspace*{0.05\linewidth}
    \begin{subfigure}{0.45\linewidth}
    \includegraphics[width=\linewidth]{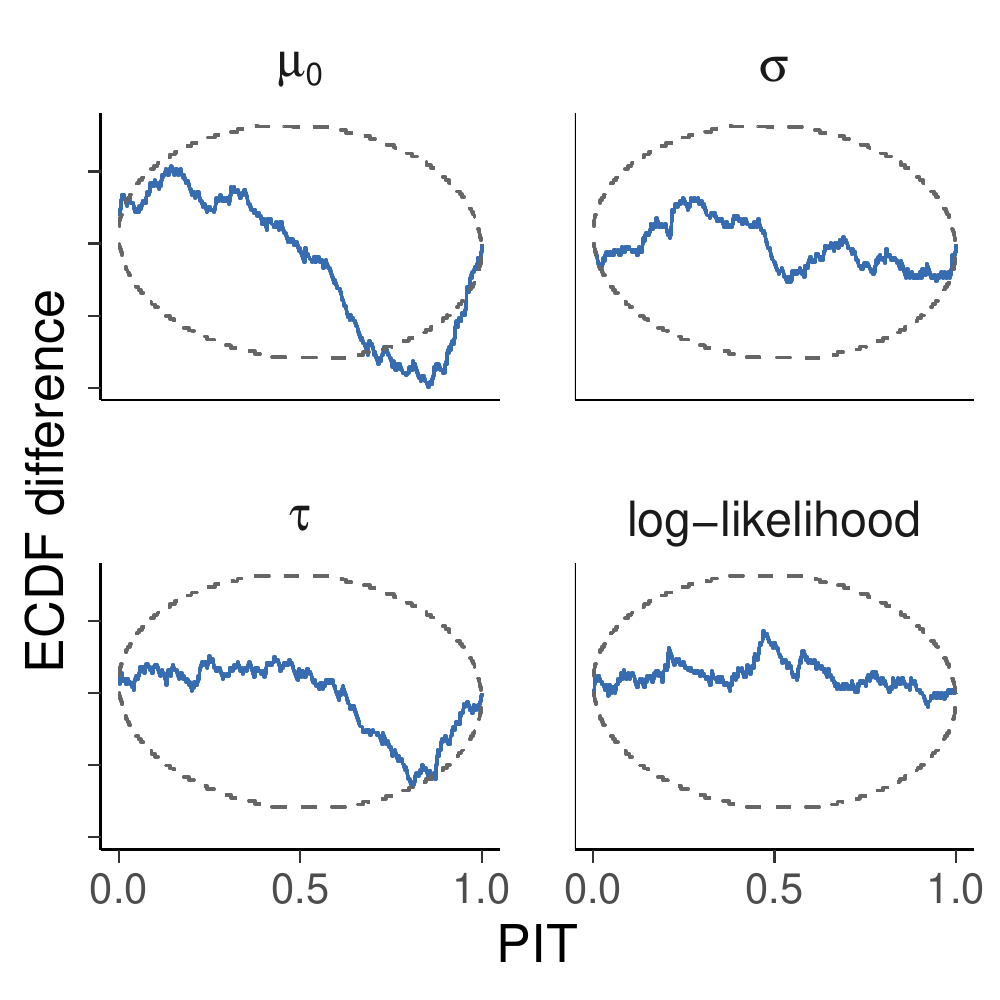}
    \caption{Non-centered Parameterization}
    \end{subfigure}
    \caption{\textit{Posterior SBC} for the hierarchical model with an observed data with large $\tau$ and small $\sigma$, corresponding to \emph{weak population prior and strong likelihood}, \new{implying a close to normal posterior with (a) the centered parameterization, and a funnel shaped posterior with (b) the non-centered parameterization. Subplots show PIT-ECDF difference plots using four different test quantities and 95\% simultaneous confidence intervals under the assumption of uniformity. The blue PIT-ECDF difference lines for the centered parameterization stay inside the envelope, indicating calibrated inference. The blue PIT-ECDF difference lines for the non-centered parametrization show that the right tail of the approximated posterior for $\mu_0$ and $\tau$ tends to be thin (the line dips outside of the envelope) when SBC iterations are averaged over the observed data posterior draws.}}
    \label{fig:hm-posterior-SBC}
\end{figure}
\begin{figure}
    \centering
    \begin{subfigure}{0.45\linewidth}
        \includegraphics[width=\linewidth]{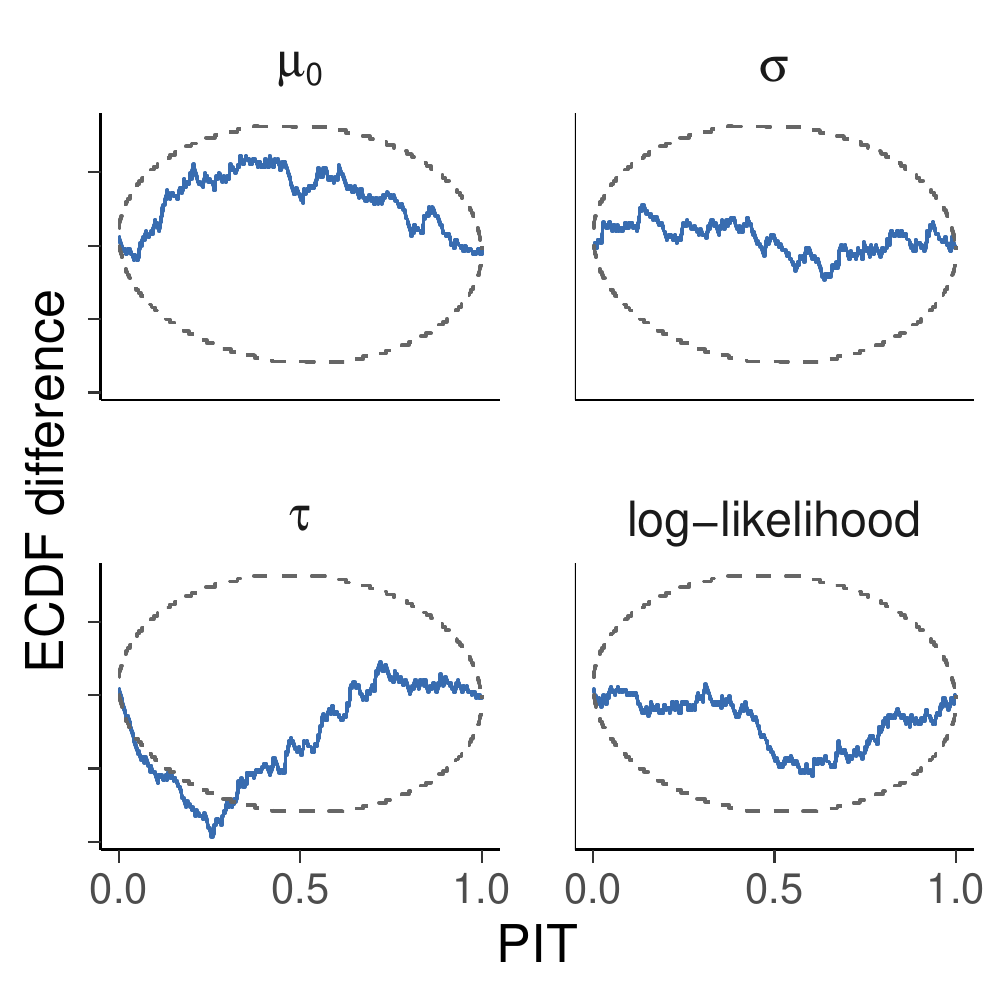}
        \caption{Centered}
    \end{subfigure}
    \hspace*{0.05\linewidth}
    \begin{subfigure}{0.45\linewidth}
    \includegraphics[width=\linewidth]{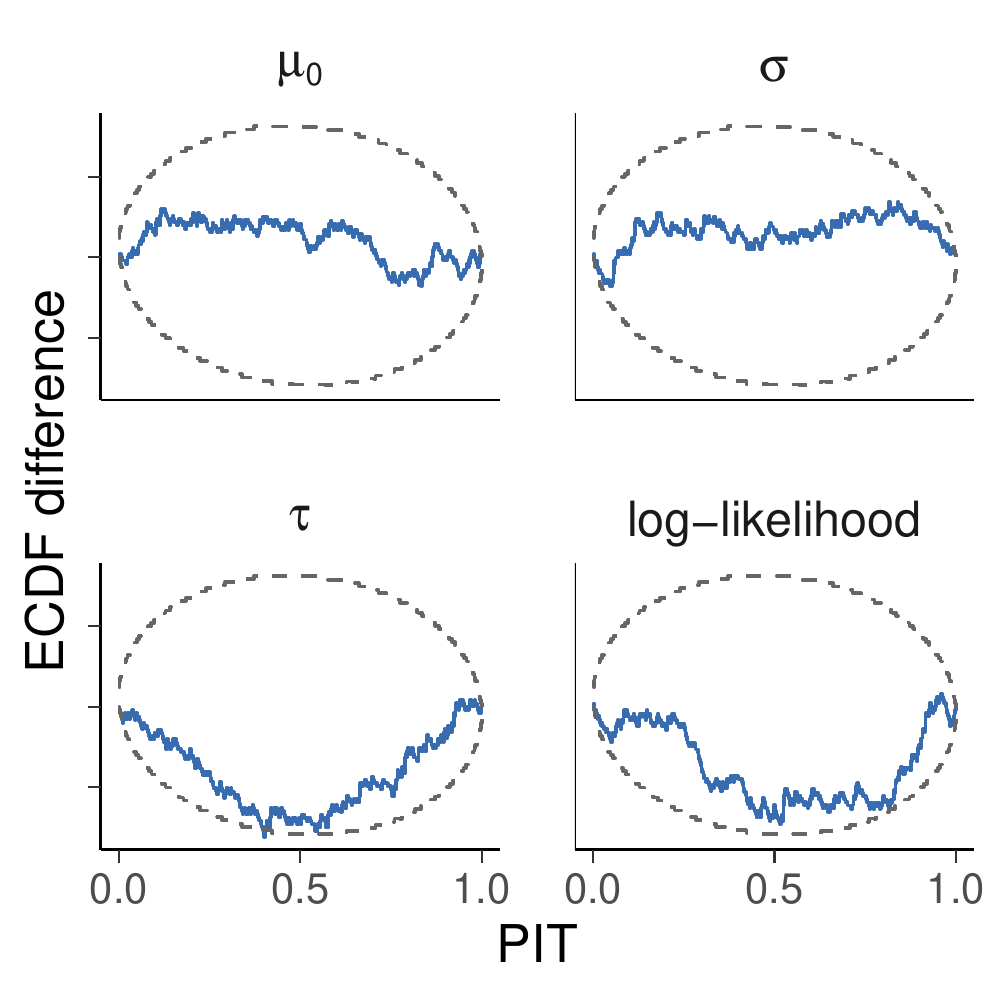}
    \caption{Non-centered}
    \end{subfigure}
    \caption{\textit{Posterior SBC} for the hierarchical model with an observed data with small $\tau$ and large $\sigma$, corresponding to \emph{strong population prior and weak likelihood}, \new{implying a funnel shaped posterior with (a) the centered parametrization, and a close to normal posterior with (b) the non-centered parametrization. Subplots show PIT-ECDF difference plots using four different test quantities and 95\% simultaneous confidence intervals under the assumption of uniformity. The blue PIT-ECDF difference lines for the centered parameterization show that the inference often overestimates $\tau$ (the line dips below the envelope), when SBC iterations are averaged over the observed data posterior draws. The blue PIT-ECDF difference lines for the non-centered parameterization stay inside the envelope, indicating calibrated inference. }}
    \label{fig:hm-posterior-SBC2}
\end{figure}

To assess the calibration of the inference, we inspect the uniformity of the PIT values of the joint log-likelihood, and of the parameters $(\mu_0$, $\tau$, $\sigma)$ shared between the groups. \new{PIT values are computed as described in Sections \ref{sec:prior_SBC} and \ref{sec: Posterior SBC}. We employ the graphical uniformity test introduced by \citet{sailynoja_graphical_2022}, see e.g., Figure~\ref{fig:hm-prior-SBC}. The test compares the empirical cumulative distribution function (ECDF) of the discrete PIT values to 95\% simultaneous central confidence bands under the assumption of discrete uniformity. We use the implementation of the test available in the \texttt{bayesplot} R package \citep{gabry_plotting_2024}. To provide better dynamic range in the plots, we show the ECDF difference (the observed ECDF minus the expected CDF values for uniform distribution) as recommended by \citet{sailynoja_graphical_2022}.} The PIT values for a given quantity are obtained by transforming the quantity with respect to the corresponding posterior draws, see Figure~\ref{fig:intro-sbc_pit}. That is, for the population parameters, the prior draw is transformed with respect to the posterior draws. For the joint log-likelihood, we compare the log-likelihood of the sample, when evaluated with the known true parameter values, against the joint log-likelihood evaluated with the parameter posterior draws. 

After prior SBC with 500 iterations, Figure~\ref{fig:hm-prior-SBC} shows no noticeable calibration issues in the inference with either parameterization. This would indicate to the modeller that everything is fine with either of these model implementations.

Next, we assess the calibration via posterior SBC by conditioning on the dataset with large $\tau$ and small $\sigma$. This corresponds to the weak population prior and strong likelihood case, implying a funnel shaped posterior with the non-centered parameterization. Running 500 iterations of posterior SBC reveals a very different story for the calibration of the two parameterizations. As shown in Figure~\ref{fig:hm-posterior-SBC}, the centered parameterization shows excellent calibration, but the non-centered parameterization has calibration issues when evaluating the population level parameters $\mu_0$ and $\tau$.

The second dataset has been generated with small $\tau$ and large $\sigma$.
This corresponds to a case of strong population prior and weak likelihood, implying a funnel shape for the posterior of the parameters in the model with centered parameterization. The posterior exhibits highly varying curvature, making it hard to explore for the inference algorithm. When we run posterior SBC conditional on this data, the presence of calibration issues with the centered parameterization of the model is confirmed. Figure~\ref{fig:hm-posterior-SBC2} shows the calibration assessment of posterior SBC, revealing the clear calibration issues with the centered parameterization, but also showing possible overestimation of $\tau$ in the inference run with the non-centered parameterization. If we would like to inspect this effect more closely, we could run posterior SBC with more iterations to increase the sensitivity of the calibration checking.

To conclude, in this example, we showcased a model where prior SBC indicates no calibration issues with either of the candidate model implementations. In contrast, with posterior SBC, we observed that for both of the inspected datasets better calibration is achieved with one of the parameterizations, conditional on a given observed dataset.

In the presented cases, the Stan interface conducted automated inference diagnostics \new{(partially in Stan itself and partially using the \texttt{posterior} R package, \citealp{posterior:2024})}, producing warnings for one or both of the parameterizations. In prior SBC both of the model parameterizations faced a considerable number of iterations with high $\widehat{R}$ values \citep{vehtari_rank-normalization_2021}, implying bad mixing of the MCMC chains. Additionally, $12\%$ of the iterations of the centered parameterization model had one or more divergent transitions, implying possibly biased posterior inference.
In posterior SBC, the first dataset with strong likelihood produced warnings for the non-centered parameterization, and the second dataset with weak likelihood was problematic for the centered parameterization.
These warnings align with our findings with SBC by also indicating the inference problems. This is plausible since the issues are caused by the challenging posterior geometry.
\new{Although in this specific case the convergence diagnostics did indicate issues, the $\widehat{R}$ and divergence diagnostics do not quantify the direction or magnitude of bias for each parameter, which on the other hand can easily be read from the graphical PIT plots. For example, Figure~\ref{fig:hm-posterior-SBC2} shows that in the centered parameterization case the inference tend to overestimate $\tau$ (dip in the left hand side of PIT ECDF-difference plot indicates missing posterior mass for small values of $\tau$), but the inference for $\sigma$ is reasonably calibrated.
Furthermore, good diagnostics are not necessarily available for new inference methods, which is demonstrated with amortized Bayesian inference in Section~\ref{sec:case-study-abi}. Finally, when the model complexity grows, the inference issues might be more subtle, but still have a strong impact on the posterior estimates and calibration.}

\subsection{Lotka-Volterra model -- focusing computational efforts}\label{sec:case-study-lv}

In this case study, we highlight how weakly informative priors can cause prior SBC to face computational issues, and how posterior SBC may provide substantial speed-ups for the calibration checking process. Additionally, we demonstrate a case where prior SBC indicates potential calibration issues for the model. However, these issues are not present in the parameter space explored by posterior SBC. Below, we show the results of performing both prior and posterior SBC for an inference task involving the Lotka-Volterra predator-prey model. We compare the findings of the calibration checking approaches, as well as the required computational efforts.

\paragraph{Setup}
The Lotka-Volterra predator-prey model consist of a system of first order non-linear differential equations, and describes the population dynamics of an interacting pair of species, a predator and its prey.
Here we model the number of snowshoe hare and Canada lynx pelts collected by the Hudson Bay Company between years 1900 and 1920 \citep{odum_fundamentals_2005}.
The expectations of the number of pelts collected are modelled as the solutions to the Lotka-Volterra equations, 
\begin{align}\label{eq:lotka-volterra-ode}
    \frac{d}{dt}H &=\alpha H - \beta H L,\\
    \frac{d}{dt}L &= -\gamma L + \delta H L,
\end{align}
where $H$ and $L$ are the populations of hares and lynxes respectively, $\alpha$ is the rate at which the hare population would grow if no lynxes were present, and conversely $\gamma$ is the death rate of lynxes without hares to hunt. The parameters $\beta$ and $\delta$ characterize the effect that the interaction between the species has on their populations -- some hares get eaten, and the lynx population may grow. The signs in the equation system are chosen so that we expect each parameter to be positive.  

We add an error term to model the measurement noise and the variation unexplained by the deterministic differential equation model. For both populations, the number of collected pelts is assumed to be log-normally distributed as
\begin{align}
    \log\left(\hat H_t\right) = \mathcal{N}\left(\log(H_t),\sigma_h\right),\\
    \log\left(\hat L_t\right) = \mathcal{N}\left(\log(L_t),\sigma_l\right),
\end{align}
where $\hat H_t$ and $\hat L_t$ are the recorded numbers of pelts collected, $H_t$ and $L_t$ the latent true (trappable) populations of the species at that time, and $\sigma_h$ and $\sigma_l$ the error terms related to the species.
All of these parameters are constrained to be positive. We do not have strong intuition on the interactions between the Lotka-Volterra model parameters, so we assign them the following independent priors, used for example in the case study of \cite{carpenter_predator-prey_2018},
\begin{align}
    \alpha, \gamma &\sim \mathcal{N}(1,0.5)\\
    \beta, \delta &\sim \mathcal{N}(0.05, 0.05). 
\end{align}
These priors are set with the aim to inform on the magnitude of the parameters and to keep the year-to-year variation of the populations from being too extreme. 
For both of the measurement error terms, we set a weakly informative log-normal prior:
\begin{align}
    \log \left(\sigma_i\right) \sim \mathcal{N}(-1,1).
\end{align}

\begin{figure}
    \centering
    \includegraphics[width=0.8\linewidth]{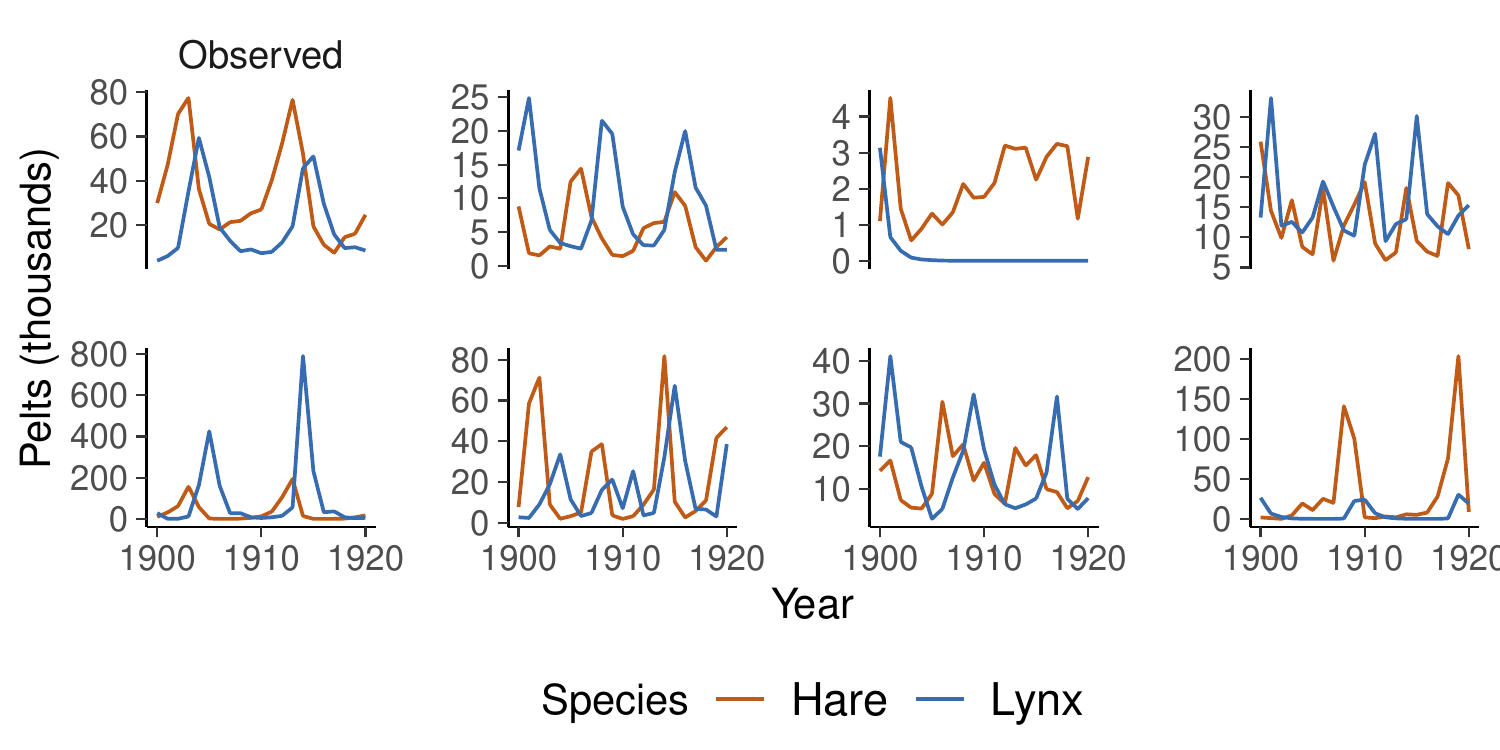}
    \caption{Prior predictive draws of the observed pelt counts over time. Top left shows the true historical observation data. Many prior draws do not exhibit the observed periodicity, or are in other ways unrealistic.}
    \label{fig:lv-prior-predictive-pelt-vs-year}
\end{figure}

Figure~\ref{fig:lv-prior-predictive-pelt-vs-year} shows some prior predictive draws from the model together with the historical observations that will later use for posterior SBC. We describe these prior predictions more below, when discussing the results of the case study. 

\begin{figure}
    \centering
    \includegraphics[width=.7\linewidth]{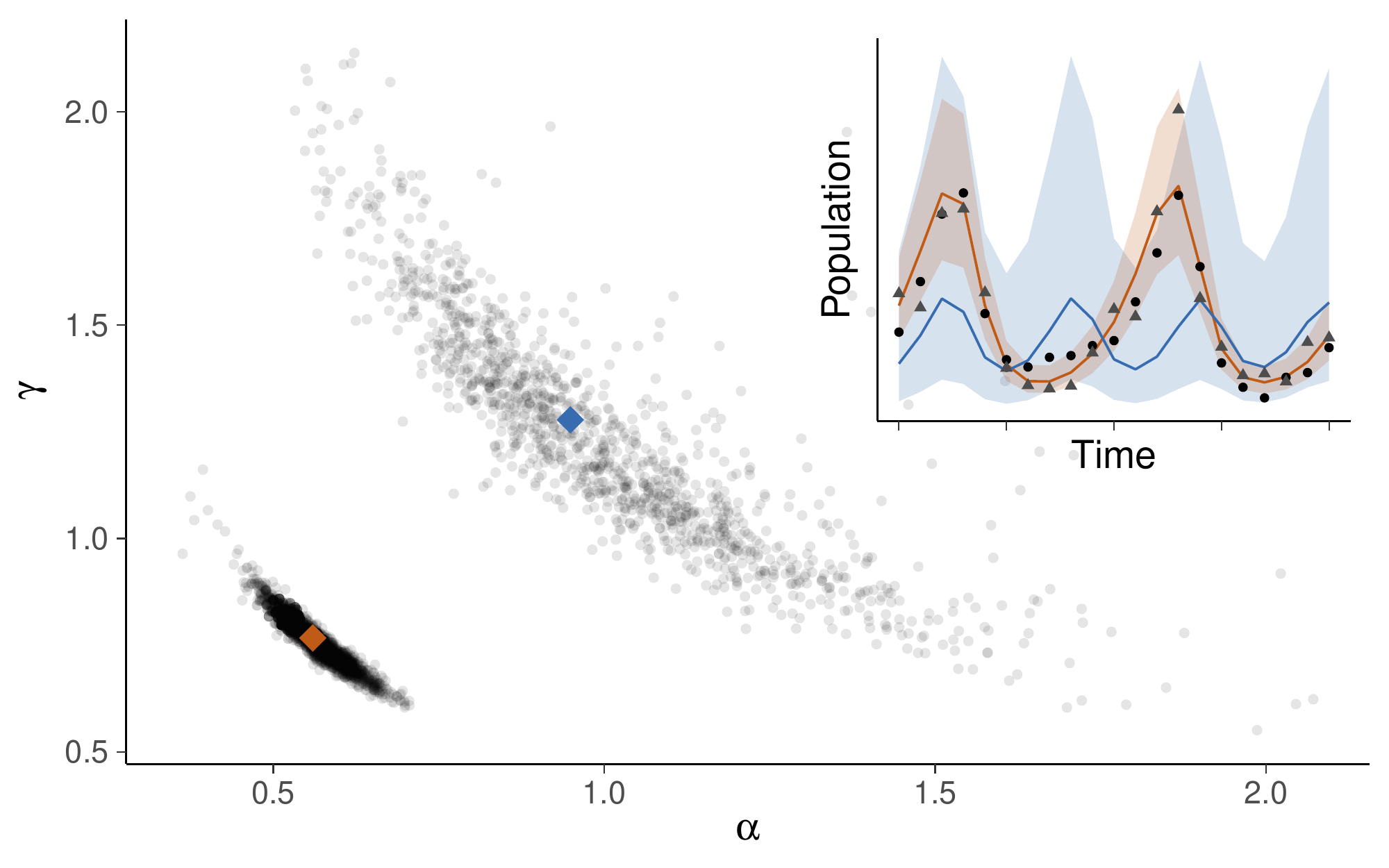}
    \caption{Bimodal posterior of the two population growth parameters, inlaid with predictive means and predictive intervals of two posterior draws where ($\theta_1$, $\theta_3$) fall in two different modes. The concentrated mode has most of the posterior mass, and corresponds to lower predictive uncertainty, and a periodicity following the data closely. The more dispersed mode, having very small amount of the posterior mass, corresponds to high observation noise.}
    \label{fig:lv-bimodal-posterior}
\end{figure}

We implement the models using Stan and use NUTS for the posterior inference. This time, we initialize the Markov chains using the Pathfinder variational inference method  \citep{zhang_pathfinder_2022}, which allows us to quickly obtain approximate draws close to the typical set of the posterior. This initialization step was added to the inference as the resulting posteriors are often multi-modal with a narrow high-probability mode corresponding to lower measurement error and a more informative likelihood for the ODE parameters, and a wider low-probability mode corresponding to high measurement error and low information on the ODE model parameters. With random initialization, some of the Markov-chains are prone to getting stuck on the low-probability mode, but Pathfinder can quickly find both modes and generate draws from the high-probability mode to be used as initial values for the Markov chains. Figure~\ref{fig:lv-bimodal-posterior} shows an example of this posterior multimodality. 

\paragraph{Results}
We set out to run 250 iterations of both prior SBC and posterior SBC, but could extend posterior SBC to 500 iterations, as the runtime per iteration was over five times faster for posterior SBC. The process of running 250 iterations of prior SBC took 6.5 hours when using parallel processing for sampling the individual MCMC chains, but sequential model fitting. Conversely, after using 10 seconds to run the initial posterior inference on the historical pelt dataset published by \cite{mahaffy_math_2010}, running 500 iterations of posterior SBC took only 2.5 hours.

\begin{figure}
    \centering
    \includegraphics[width=.6\linewidth]{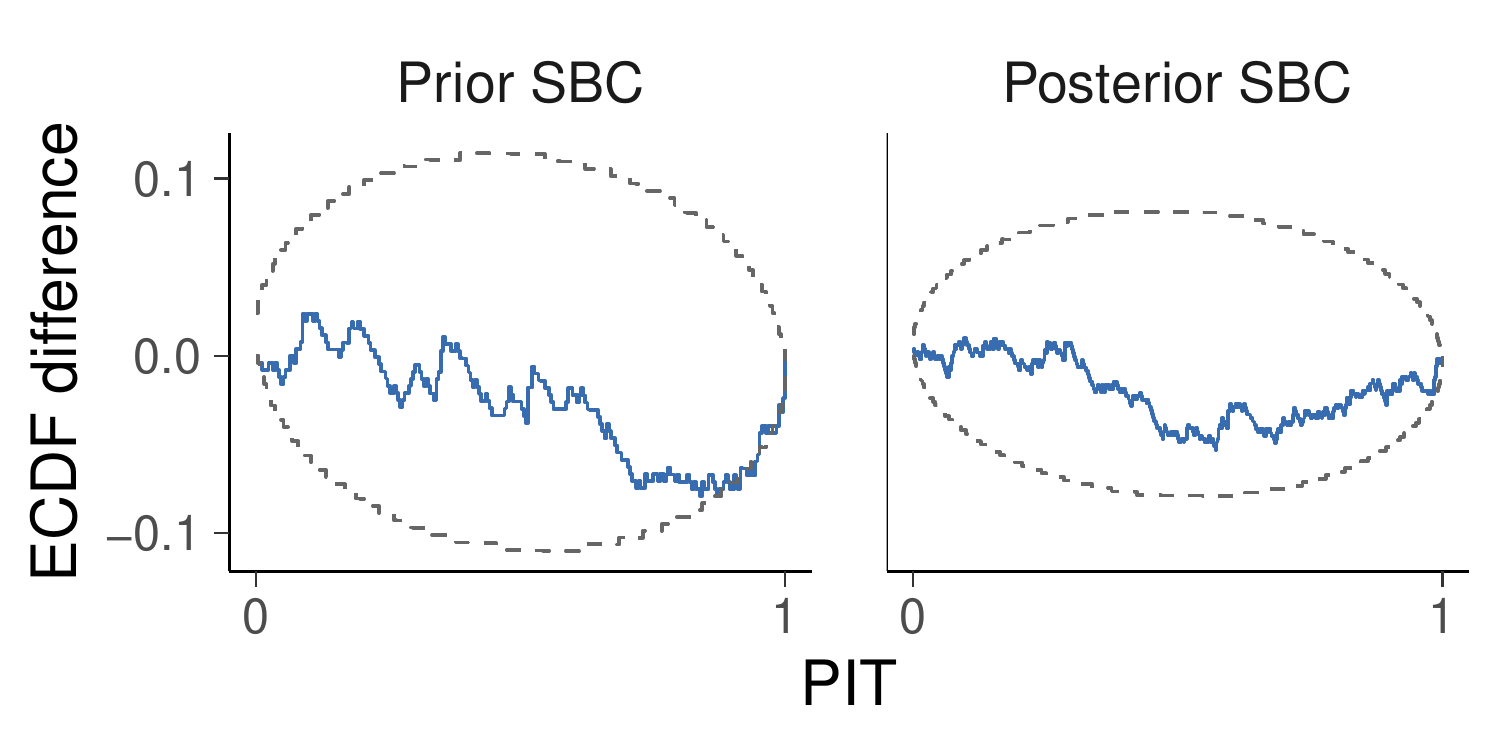}
    \caption{PIT-ECDF plots of the posterior joint log-likelihood test quantity in the Lotka-Volterra model for both prior and posterior SBC. Prior SBC uses only 250 iterations, resulting in wider confidence bands. Posterior SBC could include 500 iterations due to faster computation. \new{The blue PIT-ECDF difference line for the prior SBC dips below the envelope, indicating the inference is not well calibrated when averaged over the prior draws. The blue PIT-ECDF difference line for the posterior SBC stays inside the envelope, indicating the inference is well calibrated when averaged over the observed data posterior draws.}}
    \label{fig:lv-sbc-results}
\end{figure}

In prior SBC, we observe some prior predictive draws resulting in posterior inference with severe convergence issues. When investigating the calibration of the inference, we look at the PIT-ECDF of the joint log-likelihood in Figure~\ref{fig:lv-sbc-results}. There, we see some potential calibration issues manifesting as fewer than expected PIT values between 0.5 and 1. The calibration assessment for the individual parameters can be found in the supplementary material, but shows no calibration issues for the individual parameters before corrections for multiple testing. 

Posterior SBC, in turn, indicates good calibration for the observed historical data. A graphical assessment of the uniformity of the PIT joint log-likelihood values is shown in Figure~\ref{fig:lv-sbc-results}. The faster inference during the posterior SBC allowed us to iterate on the model and experiment on using Pathfinder for initializing the MCMC chains in the more desirable posterior mode. More details are shown in the supplementary material at Appendix~A.

To conclude, we have demonstrated a case, where the calibration issues indicated by prior SBC are not present once we condition on an observed dataset. Moreover, due to the problematic inference in prior SBC, posterior SBC turns out to be computationally less demanding and thus allows us to obtain more power on the calibration assessment while using a lower computational budget.

The observed inference problems during prior SBC are due to the independent weakly informative priors assigning a considerable prior probability to problematic parameter combinations, that are no longer plausible under the observation. Furthermore, when we inspect the prior predictive samples in ~Figure~\ref{fig:lv-prior-predictive-pelt-vs-year}, we observe multiple samples that do not align with our expectations on the periodic nature of the data. We could try to improve the prior, but if the computation already works for the posterior it is not needed.

\new{\citet{obrien+etal:2025:numerical_integration_error} demonstrate a case where numerical error in the ODE solver distorts the posterior. From the usual inference diagnostic perspective, the bias due to the distortion is not distinguishable from the true posterior as the numerical error alters the target distribution itself. However, the distortion from the numerical error is unlikely to exhibit consistency over Bayesian updating, and thus posterior SBC would indicate the problem.}

\FloatBarrier

\subsection{Posterior SBC in amortized Bayesian inference}\label{sec:case-study-abi}

In this case study, we show how posterior SBC can serve as a powerful diagnostic for amortized Bayesian inference workflows.
Amortized Bayesian inference \citep[see][for an overview]{zammit-mangion_abi_overview_2024} constitutes a new wave of Bayesian inference methods, where neural networks learn a direct surrogate for the posterior distribution.
Normalizing flows \citep{papamakarios2021} are arguably the most prominent neural density estimator for amortized inference, but other architectures have recently been explored as well, such as score-based diffusion models \citep{geffner2022diffusionsbi,sharrock2022diffusionsbi}, flow matching \citep{dax2023fmpe}, and consistency models \citep{schmitt2023cmpe}. \new{For amortized inference, there are no inference diagnostics similar to the convergence diagnostics for MCMC. While posterior SBC for MCMC has the high cost of running MCMC many times, in amortized inference the additional cost of running posterior SBC is negligible.}

\begin{figure}[t]
    \centering
    \includegraphics[width=1.0\linewidth]{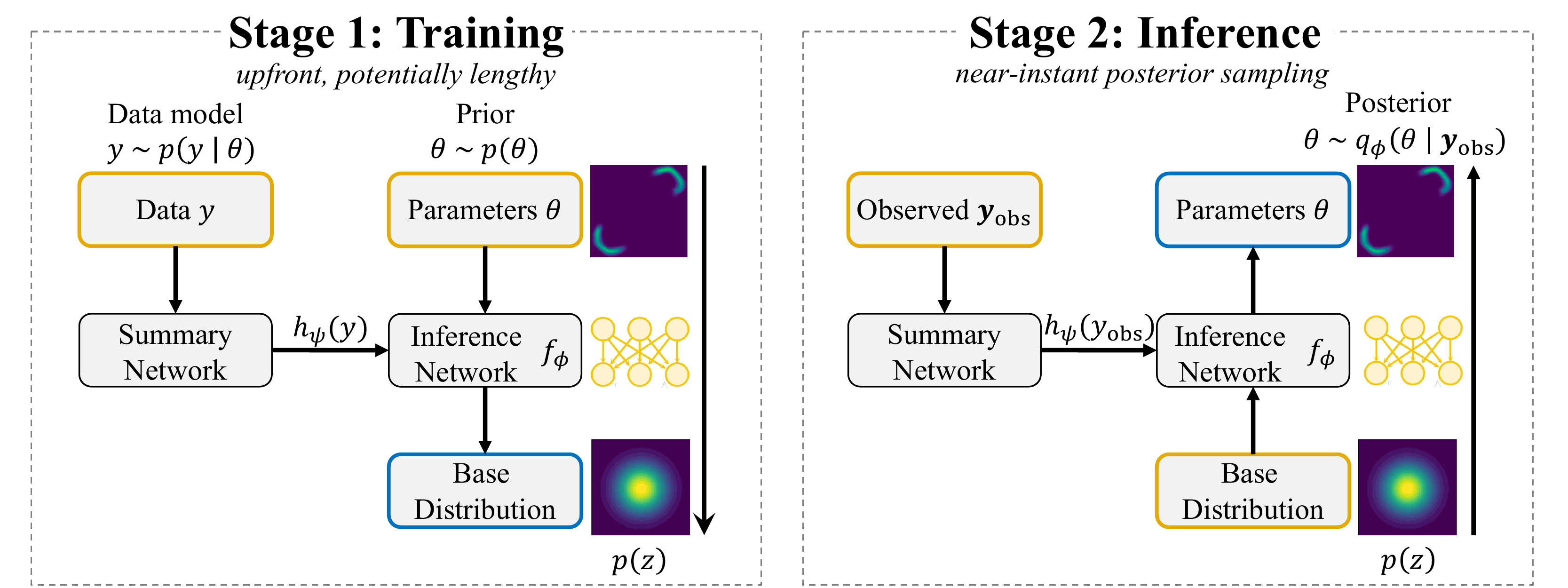}
    \caption{
    Conceptual overview of amortized Bayesian inference with normalizing flows.
    \textbf{Stage 1: Training.} Based on samples from the joint model $p(\theta, y)$, a neural network tandem simultaneously learns to extract sufficient summary statistics $h_{\psi}(y)$ and establish a conditional mapping to a base distribution $p(z)$.
    \textbf{Stage 2: Inference.} Given a new unseen dataset $\yobs$, we can draw samples from the base distribution $p(z)$ and pass them through the inverted inference network conditional on the summary statistics $h_{\psi}(\yobs)$ to obtain samples from the approximate posterior in near-instant time.
    }
    \label{fig:abi-overview}
\end{figure}

Amortized inference consists of two stages: (1) A training stage, where a generative neural network $q_{\phi}$ learns to distil relevant information from the probabilistic model based on synthetic data $(\theta, y)\sim p(\theta, y)$ from the joint model; and (2) the inference stage, where the trained neural networks approximate the posterior distribution for an arbitrary new dataset $y_{\mathrm{obs}}$.
Amortized inference casts fitting a probabilistic model as a neural network prediction task, which typically takes well under a second.
This achieves near-instant approximate posterior sampling $\theta\sim q_{\phi}(\theta\given y_{\mathrm{obs}})$ and even direct posterior density evaluations for arbitrary parameter values $\theta$.

In an effort to align amortized inference with the needs of common probabilistic modeling settings, there are two crucial extensions to the standard approach described above.
First, in Bayesian inference, the data $y$ can be replaced by \emph{sufficient} summary statistics $h^*(y)$ without altering the posterior: $p(\theta\given y) = p(\theta\given h^*(y))$.
In amortized inference, neural networks are employed to learn embeddings of the data \emph{in tandem} with the posterior approximator \citep{radev2020bayesflow, radev2020towards, chen2021neural, chan2018likelihood,huang2023,schmitt2022bayesflow}.
These \emph{summary networks} $h_{\psi}$ are parameterized by learnable neural network weights $\psi$ and learn a fixed-length representation of the data which is approximately sufficient for posterior inference (not necessarily for reconstructing the data).
Second, datasets $y = y_1, \ldots, y_N$ can come with a varying number of observations $N$.
While the summary network learns to compress datasets of varying length to a fixed-length vector $h_{\psi}(y)$, the Bayesian posterior is influenced by the number of observations $N$ (e.g., as described by contraction).
Hence, the neural network needs to be informed about the number of observations in the raw dataset, and we need to cover datasets with varying numbers of observations in the neural network training stage.
Accordingly, we draw the number of observations in each \emph{simulated} dataset from a distribution $p(N)$ and condition the generative neural network on $N$ in addition to the learned summary statistics $h_{\psi}(y)$.
In summary, the forward simulation process is formalized as
\begin{equation}
    \begin{aligned}
    N \sim p(N), \quad
    \theta \sim p(\theta), \quad
    y_{1}, \ldots, y_{N} \sim p(y\given\theta, N).
    \end{aligned}
\end{equation}

The resulting neural network training for a conditional normalizing flow with learned summary statistics $h_{\psi}$ and the number of observations $N$ as additional conditioning variable minimizes the maximum likelihood objective,
\begin{equation}
    \psi^*, \phi^* = \argmin_{\psi, \phi} \mathbb{E}_
    {p(N, \theta, y_1, \ldots, y_N)}
    \Big[-\log q_\phi\Big(\theta\,\Big|\, h_{\psi}(y_1,\ldots,y_N), N\Big)\Big],
\end{equation}
where the expectation is taken over the joint distribution $p(N, \theta, y)$.

However, there is no free lunch, and state-of-the-art amortized inference methods still lack the gold-standard guarantees from established algorithms like MCMC.
The two most pressing issues of amortized neural posterior estimators are (1) the increased number of design choices from neural network architectures and training hyperparameters; and (2) the lack of performance guarantees under domain shifts between the simulated training data and the real data $\yobs$, for example resulting from model misspecification.
As a consequence, there is a need for strong diagnostics to gauge the trustworthiness of amortized neural posterior approximators.
In the context of non-amortized MCMC, one bottleneck of simulation-based calibration (both prior and posterior SBC) is the associated computational burden from repeated model re-fits on new simulated datasets.
Yet, an \emph{amortized} approximator can sample the required approximate posterior draws for new datasets in near-instant time, which reduces the runtime for SBC to only a few seconds.
Therefore, posterior SBC naturally lends itself to amortized inference due to (1) the particular need for strong diagnostics; and (2) the stunningly quick runtime with an amortized approximator.

\paragraph{Setup}
To improve our understanding of human cognition, researchers employ increasingly sophisticated statistical models to explore the neurological connections between cognitive processes and physical phenomena.
From a mathematical perspective, human decision-making can be represented by a stochastic evidence accumulation process, specifically through a drift-diffusion model \citep[DDM;][]{Voss2004,Ratcliff2008}. 
The DDM assumes that human decision-making is based on sequential integration of evidence over time until it reaches a threshold. 
The model is parameterized by an individual's cognitive parameters (e.g., information uptake speed and decision thresholds).
In its general form, a DDM is mathematically described by a differential equation,
\begin{equation}
    \diff y_t = v\delta\diff t + \sigma W_t,
\end{equation}
where $\delta$ is the drift rate and $\sigma$ controls the stochastic variance of the Wiener process $W_t$.
Simulation programs use a discretized version of the process,
\begin{equation}
    y_{t+\Delta t} = y_{t} + \delta\Delta t + \sigma \varepsilon \sqrt{\Delta t},
\end{equation}
where $\varepsilon\sim\mathcal{N}(0, 1)$ is Gaussian noise and $\Delta t$ controls the time resolution (5 milliseconds in our data model).
The DDM is additionally parameterized by a decision threshold $\alpha$, an evidence starting point $\beta$ to represent biases, and a non-decision time $\tau$ to account for motor reaction times.
We refer to Appendix~B in the supplementary material for more details.

In a line of applied work, neurological research has identified neural markers that are associated with decision-making processes. 
In a recent paper, \citet{ghaderi-kangavari_general_2023} propose a set of multiple statistical models that integrate both the cognitive drift-diffusion model and a compressed representation of neural measurements, as found in EEG or fMRI data.
Such integrative models, which combine neural processes at the level of single experimental trials, represent the current state-of-the-art in cognitive modeling research.
This case study implements two probabilistic models $M_1$ and $M_2$, which correspond to models \#2 and \#6 from \citet{ghaderi-kangavari_general_2023}.
We use deep neural networks to fit an amortized posterior approximator for each model.
Detailed specifications of models and neural networks can be found in Appendix~B in the supplementary material.
We showcase how posterior SBC can yield important information to compare the neural approximations of the candidate models conditional on two real datasets from another study \citep{Georgie2018}.

\begin{figure}[t]
    \centering
    \begin{subfigure}{1.0\linewidth}
        \setlength{\tabcolsep}{1pt}
        \begin{tabular}{cccc}
        {\small Prior SBC, $N=\Nobs$}
        &{\small Prior SBC, $N=2\,\Nobs$}
        &{\small Posterior SBC $\yobs^{(1)}$}
        &{\small Posterior SBC $\yobs^{(2)}$}\\
        \includegraphics[width=0.24\linewidth]{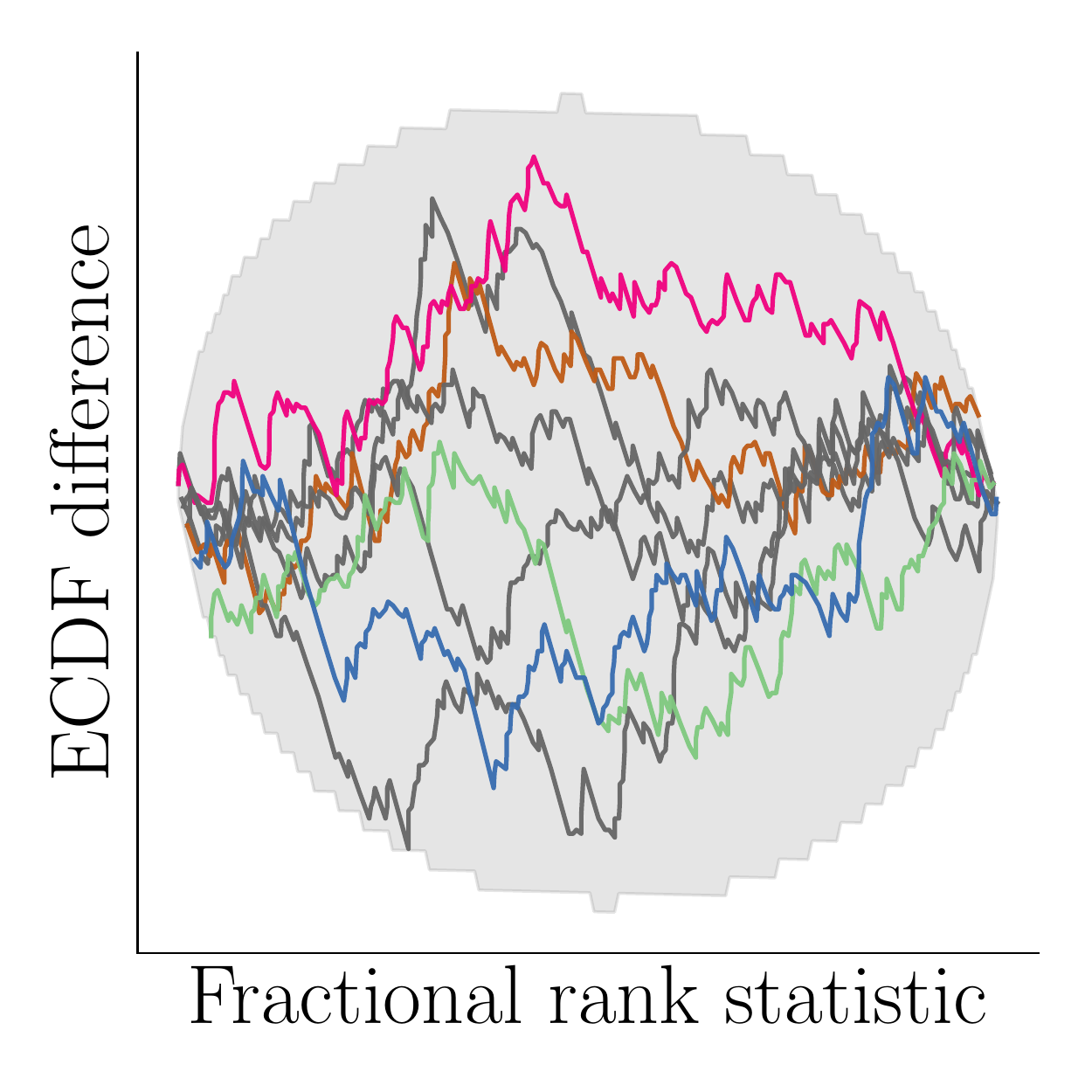}    
        &\includegraphics[width=0.24\linewidth]{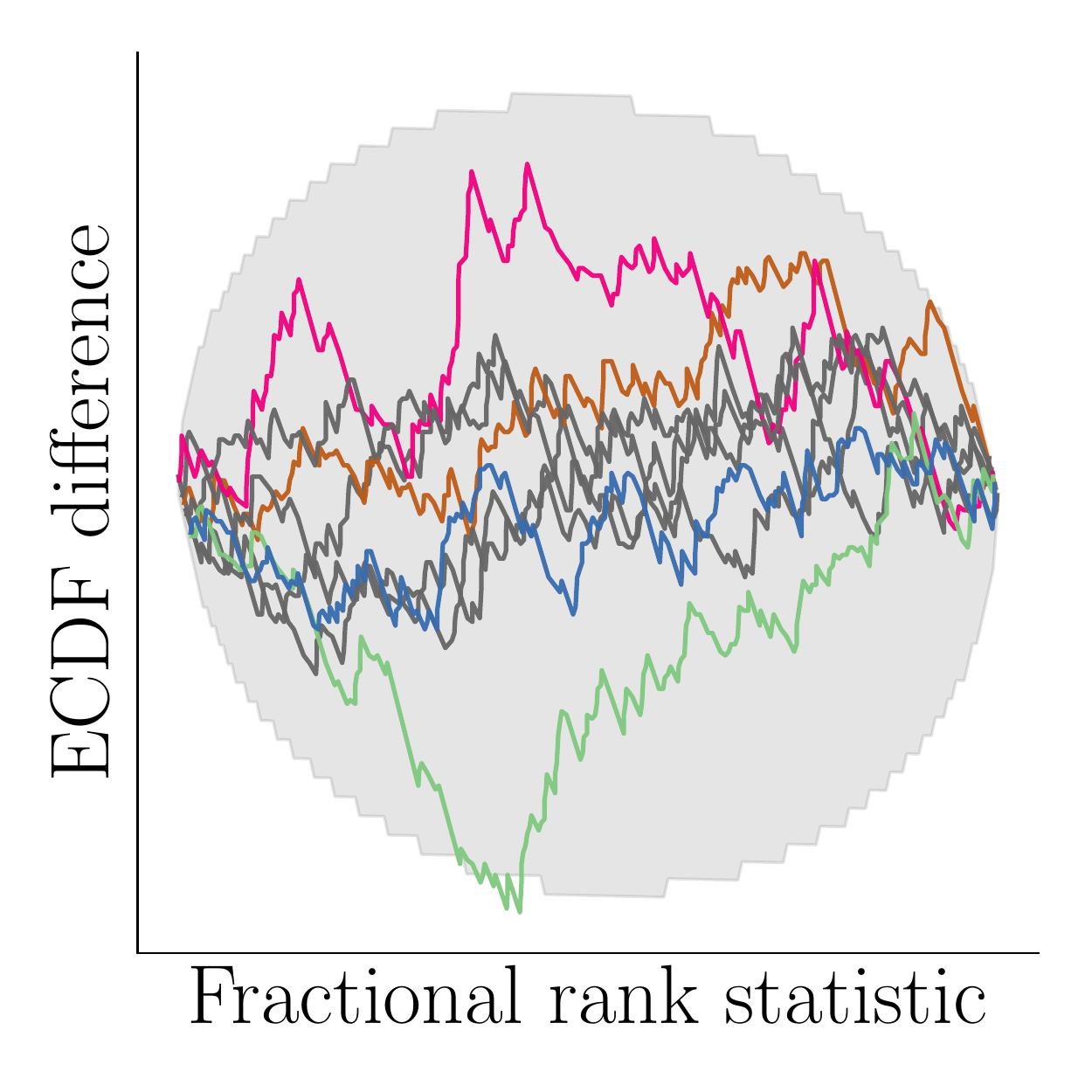}    
        &\includegraphics[width=0.24\linewidth]{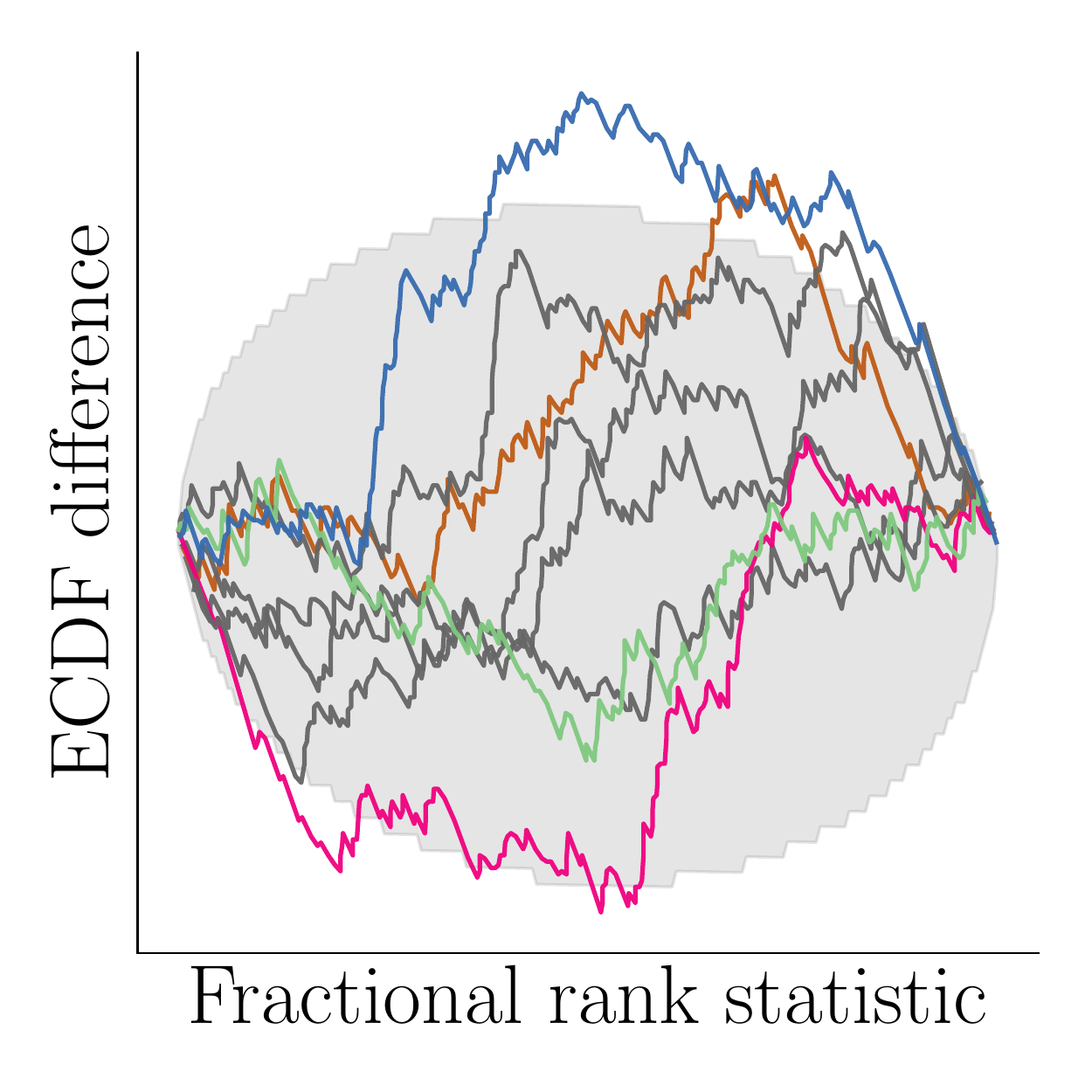}
        &\includegraphics[width=0.24\linewidth]{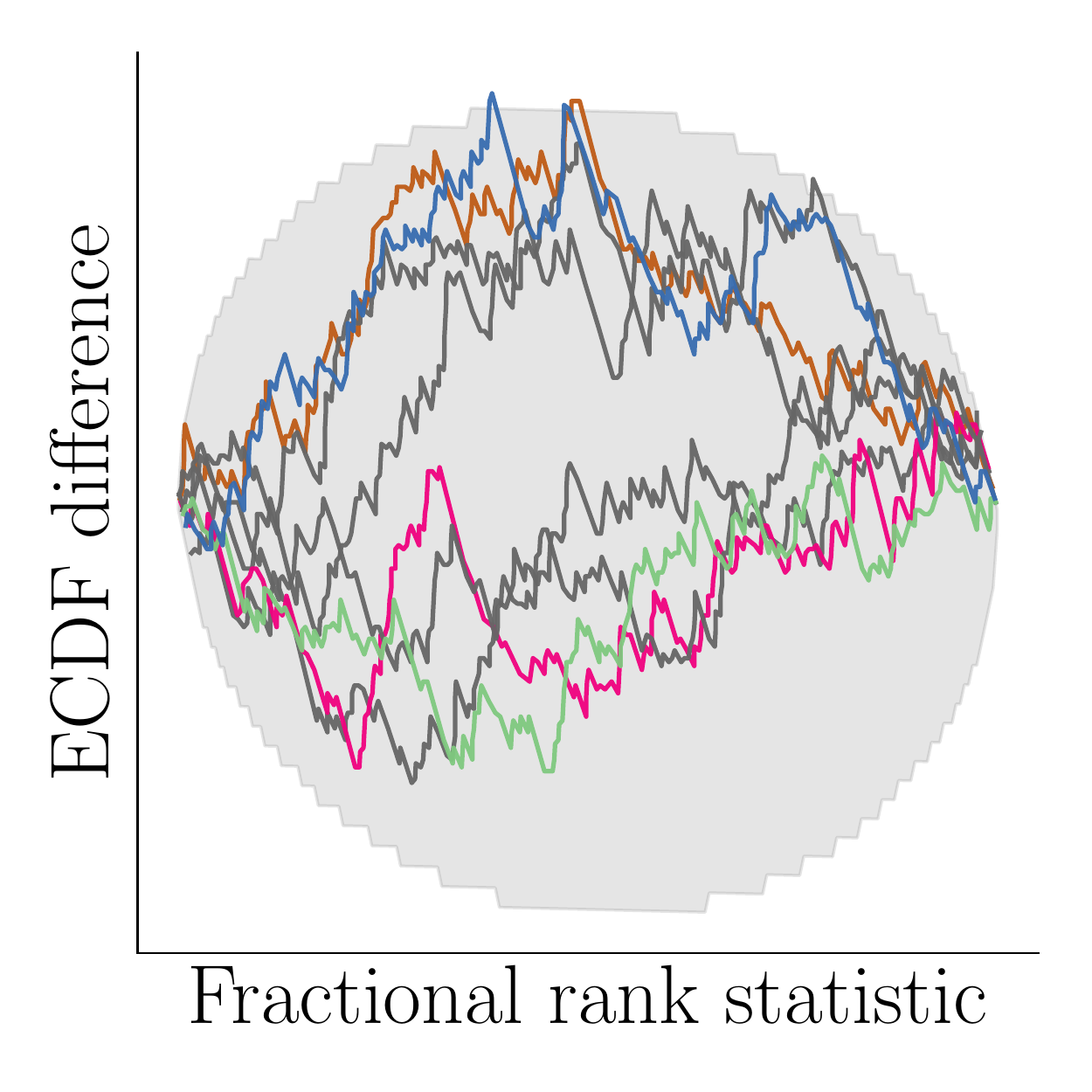}    
        \end{tabular}
        \caption{Candidate model $M_1$ with highlighted parameters
        \textcolor[HTML]{bf5b17}{$\alpha$},
        \textcolor[HTML]{f0027f}{$\mu_{(e)}$},
        \textcolor[HTML]{7fc97f}{$s_{\tau}$},
        \textcolor[HTML]{386cb0}{$\gamma$}.}
        \label{fig:abi-results:m1}
    \end{subfigure}
    \begin{subfigure}{1.0\linewidth}
        \setlength{\tabcolsep}{1pt}
        \begin{tabular}{cccc}
        {\small Prior SBC, $N=\Nobs$}
        &{\small Prior SBC, $N=2\,\Nobs$}
        &{\small Posterior SBC $\yobs^{(1)}$}
        &{\small Posterior SBC $\yobs^{(2)}$}\\
        \includegraphics[width=0.24\linewidth]{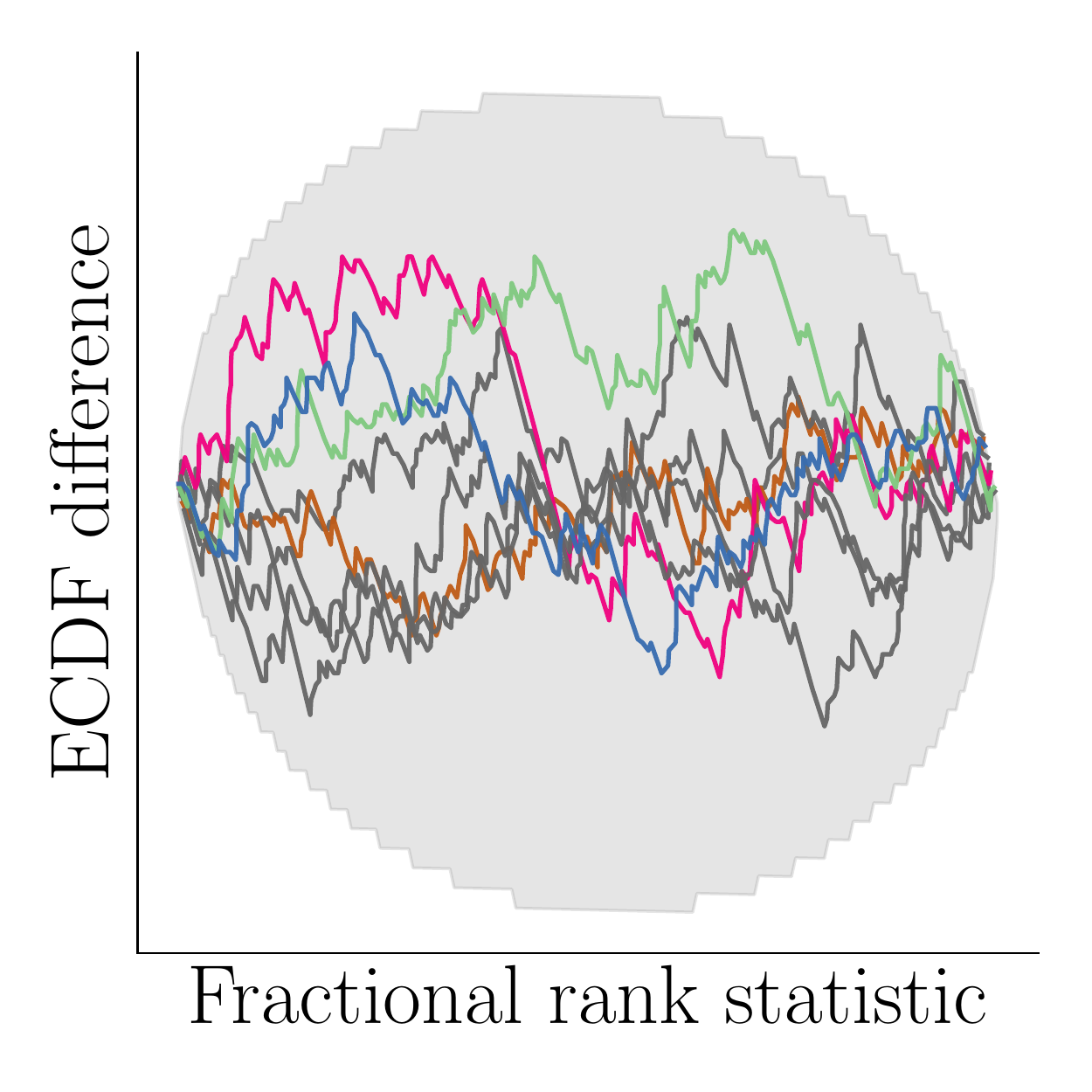}    
        &\includegraphics[width=0.24\linewidth]{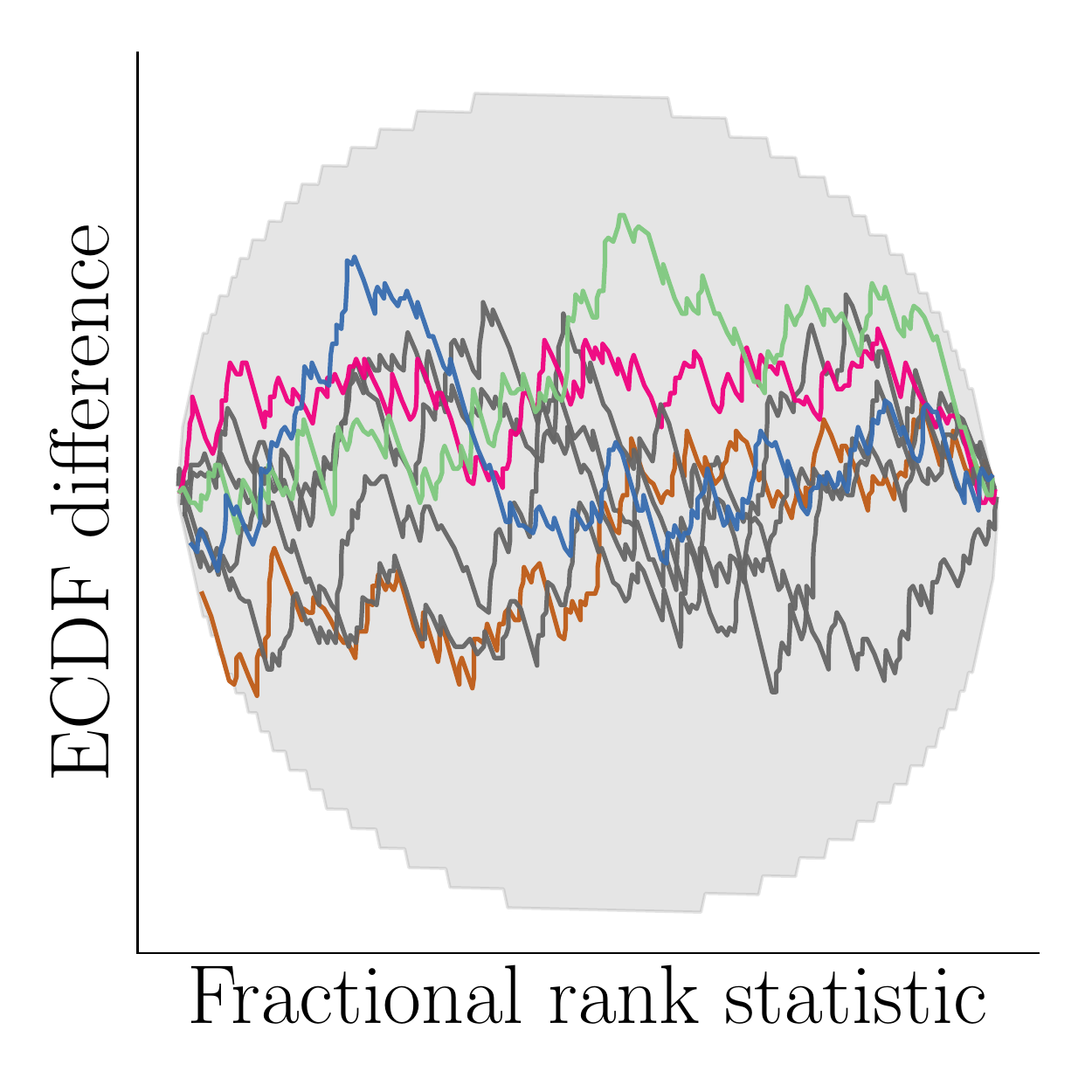}    
        &\includegraphics[width=0.24\linewidth]{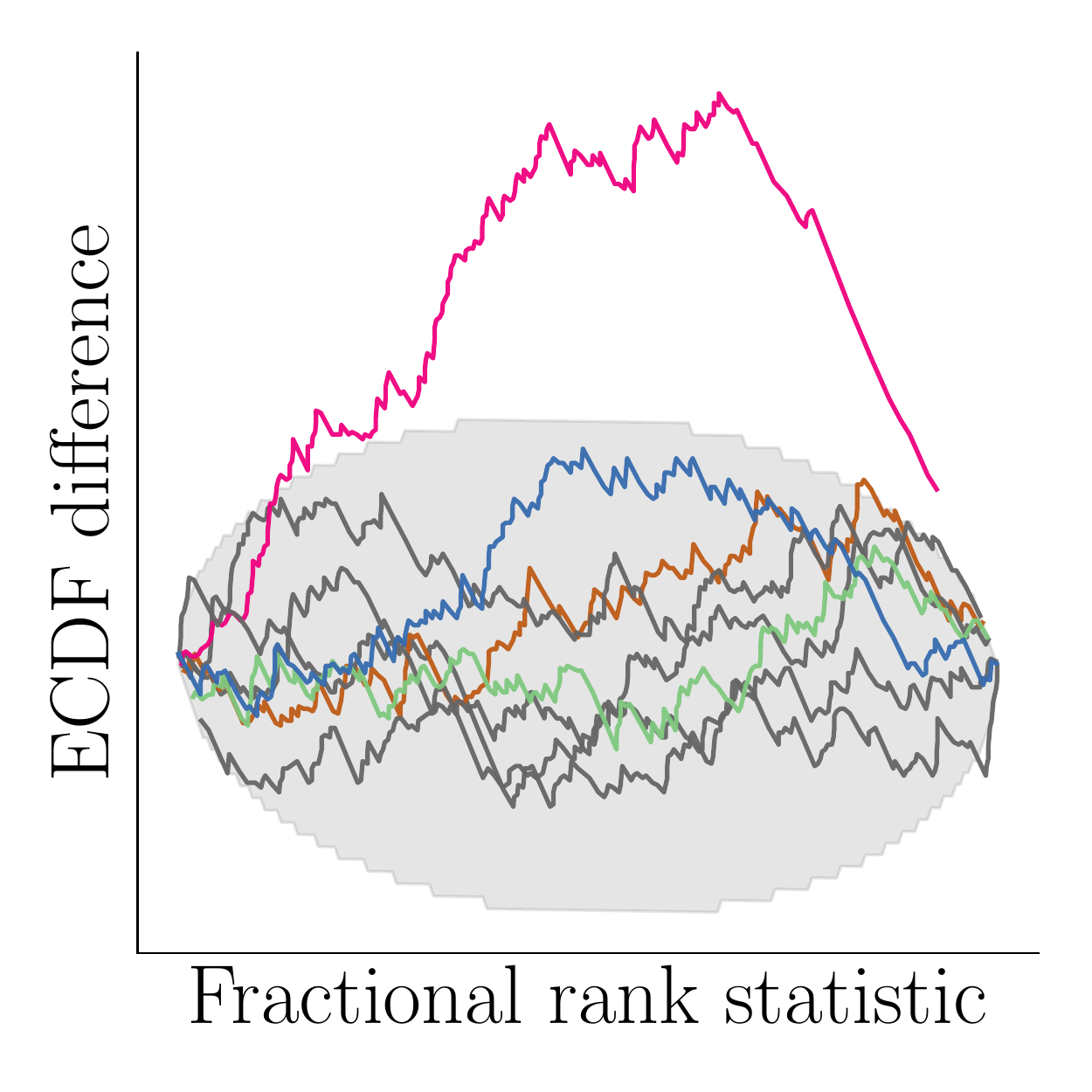}
        &\includegraphics[width=0.24\linewidth]{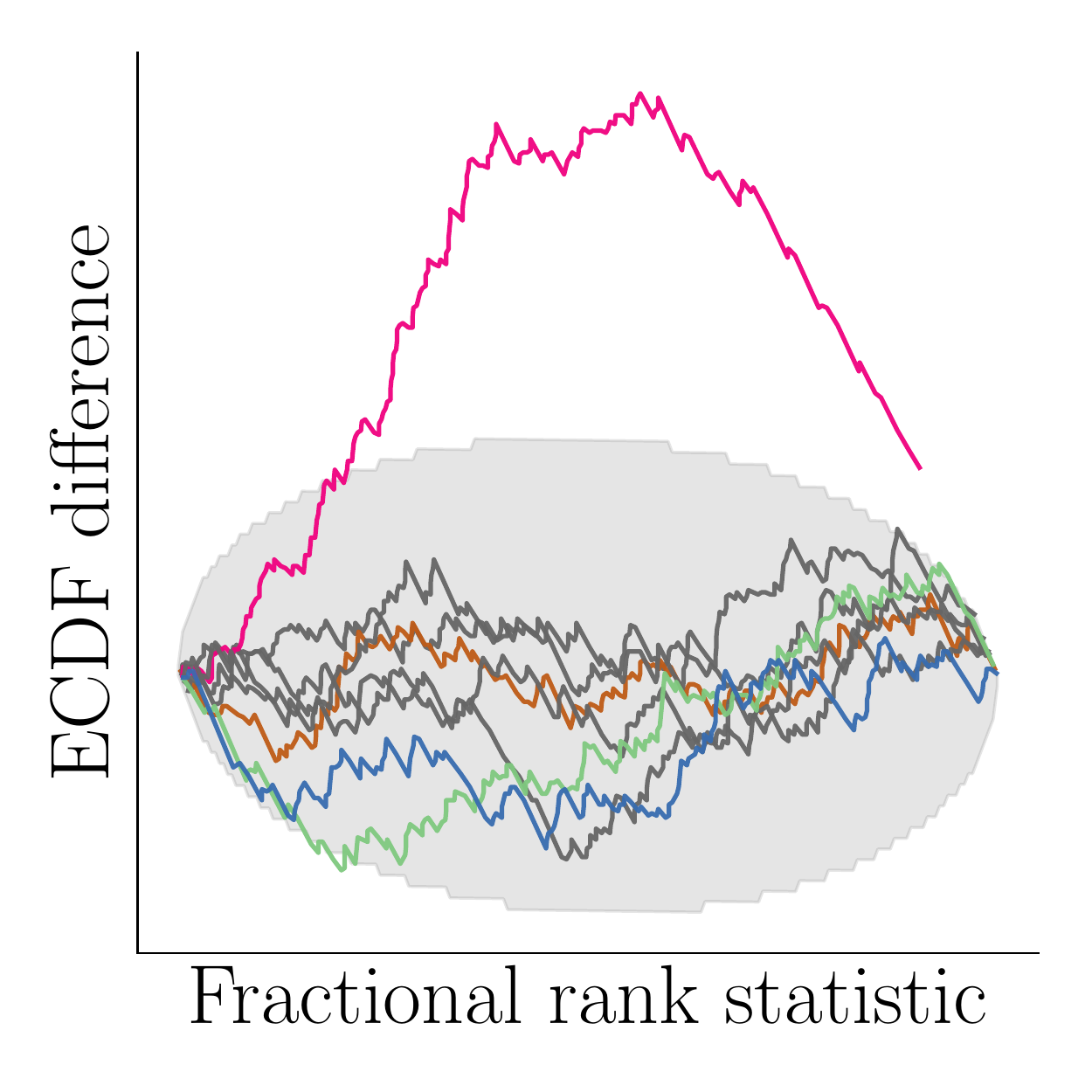}    
        \end{tabular}
        \caption{Candidate model $M_2$ with highlighted parameters
        \textcolor[HTML]{bf5b17}{$\alpha$},
        \textcolor[HTML]{f0027f}{$\mu_{(e)}$},
        \textcolor[HTML]{7fc97f}{$s_{\tau}$},
        \textcolor[HTML]{386cb0}{$\lambda$}.}
        \label{fig:abi-results:m2}
    \end{subfigure}
    \caption{Results of prior and posterior simulation-based calibration checking with amortized inference.
    The Bayesian models $M_1$ and $M_2$ are two joint integrative neuroscience models from \citet{ghaderi-kangavari_general_2023}, and the datasets for posterior SBC checking are two real datasets $\yobs^{(1)}, \yobs^{(2)}$ from \citet{Georgie2018}.}
    \label{fig:abi-results}
\end{figure}

\paragraph{Results}
In the following, we summarize the main results of the case study and refer to Appendix~A in the supplementary material for additional details.
In neural amortized inference, we must check the inference validity for different numbers of observations because the latter is passed as a conditioning variable to the inference network $f_{\phi}$.
To this end, we confirm that amortized inference on synthetic datasets simulated from the joint model $p(N,\theta,y)$ can recover the ground-truth values of the model parameters that are identifiable for both $\Nobs$ and $2\Nobs$ (see Appendix~A in the supplementary material).
Further, we can confirm that prior SBC checking for datasets with $\Nobs$ as well as $2\Nobs$ yields satisfactory results (see Figure~\ref{fig:abi-results}).
However, this conclusion drastically changes when we perform SBC checking conditional on the real datasets $\yobs^{(1)}$ and $\yobs^{(2)}$.
As shown in Figure~\ref{fig:abi-results:m1}, posterior SBC indicates minor data-conditional calibration issues of model $M_1$ for the first observed dataset $\yobs^{(1)}$ while calibration conditional on the second observed dataset $\yobs^{(2)}$ is within the acceptance band.
The second model $M_2$, however, shows pathologically bad calibration of the mean visual encoding latency $\mu_{(e)}$ conditional on either observed dataset (see Figure~\ref{fig:abi-results:m2}).
As a consequence, model $M_2$ might be refined according to the iterative Bayesian workflow \citep{Gelman_bayesian_2020,schad_toward_2020} with a particular focus on the mean visual encoding latency parameter $\mu_{(e)}$.

In this case study, posterior SBC could flag calibration issues of an amortized approximator that were not detectable with prior SBC, which illustrates its potential as a default data-conditional diagnostic in amortized Bayesian workflows \citep{radev2023BayesFlowAmortizedBayesian,schmitt2024AmortizedBayesianWorkflow}.

\section{Discussion}\label{sec: conclusions}

\new{Prior SBC and posterior SBC have different roles in the Bayesian workflow. Posterior SBC can be used to diagnose whether the used inference algorithm is well calibrated, that is, sampling faithfully from the posterior and the augmented posteriors, which are close to the original posterior.  In case of miscalibration, posterior SBC can indicate also the direction and magnitude of the bias for each parameter or other quantity of interest.}
If there is a worry that the augmented posterior having double data size is too different from the original posterior, it is possible to only use part of the data for the first posterior and make the augmented data size more similar to the original data size.

Passing posterior 
\new{SBC checking is a necessary but not sufficient condition to trust the posterior inference in the same way as most convergence diagnostics given finite computation time. If the common MCMC diagnostics, like $\widehat{R}$ and divergences,indicate big problems, it is good to first investigate the possible reasons before spending computation time for posterior SBC. Even if the convergence issues are not completely solved, it is possible to use posterior SBC to assess the direction and magnitude of the potential bias.}

Posterior SBC is specifically useful when the usual diagnostics may have missed something, when the usual diagnostics may have false positive warnings, when using new algorithm implementations that are not yet trusted, or when using new inference algorithms which do not yet have faster diagnostics. Posterior SBC seems to be specifically useful for amortized inference, as the computational cost for repeated inference is low.

\subsection*{Data and code availability}

Code to reproduce all experiments is available in the public repository at \url{https://github.com/TeemuSailynoja/posterior-sbc}.

\subsection*{Acknowledgments}
We thank Martin Modrák, Valentin Pratz, and Lasse Elsemüller for helpful comments and suggestions, and we acknowledge the computational resources provided by the Aalto Science-IT project via the \emph{Triton} compute cluster.
TS and AV acknowledge the support by the Research Council of Finland Flagship programme: Finnish Center for Artificial Intelligence, and  Research Council of Finland project (340721).
MS and PCB acknowledge support of Cyber Valley Project CyVy-RF- 2021-16 and the DFG under Germany’s Excellence Strategy – EXC-2075 - 390740016 (the Stuttgart Cluster of Excellence SimTech).
MS acknowledges the European Union’s Horizon 2020 research and innovation programme under grant agreements No 951847 (ELISE) and No 101070617 (ELSA), and the ELLIS PhD program. PCB acknowledges support of the DFG Collaborative Research Center 391 (Spatio-Temporal Statistics for the Transition of Energy and Transport) -- 520388526.

\afterpage{\FloatBarrier}
\bibliography{references2,references_abi,references_workflow}

\clearpage
\appendix
This supplement provides details and additional results for the case studies of the main paper. The source code for reproducing these case studies is available at \url{https://github.com/TeemuSailynoja/posterior-sbc}.

\subsection*{Appendix A: Case study 2: Lotka-Volterra model}\label{sec:app:lv}

Below, we present calibration assessments, as well as parameter recovery results for the individual model parameters for both prior and posterior SBC of the Lotka-Volterra model implementation.

\begin{figure}[p]
    \centering
    \begin{subfigure}[b]{0.9\linewidth}
    \centering\includegraphics[width=.8\linewidth]{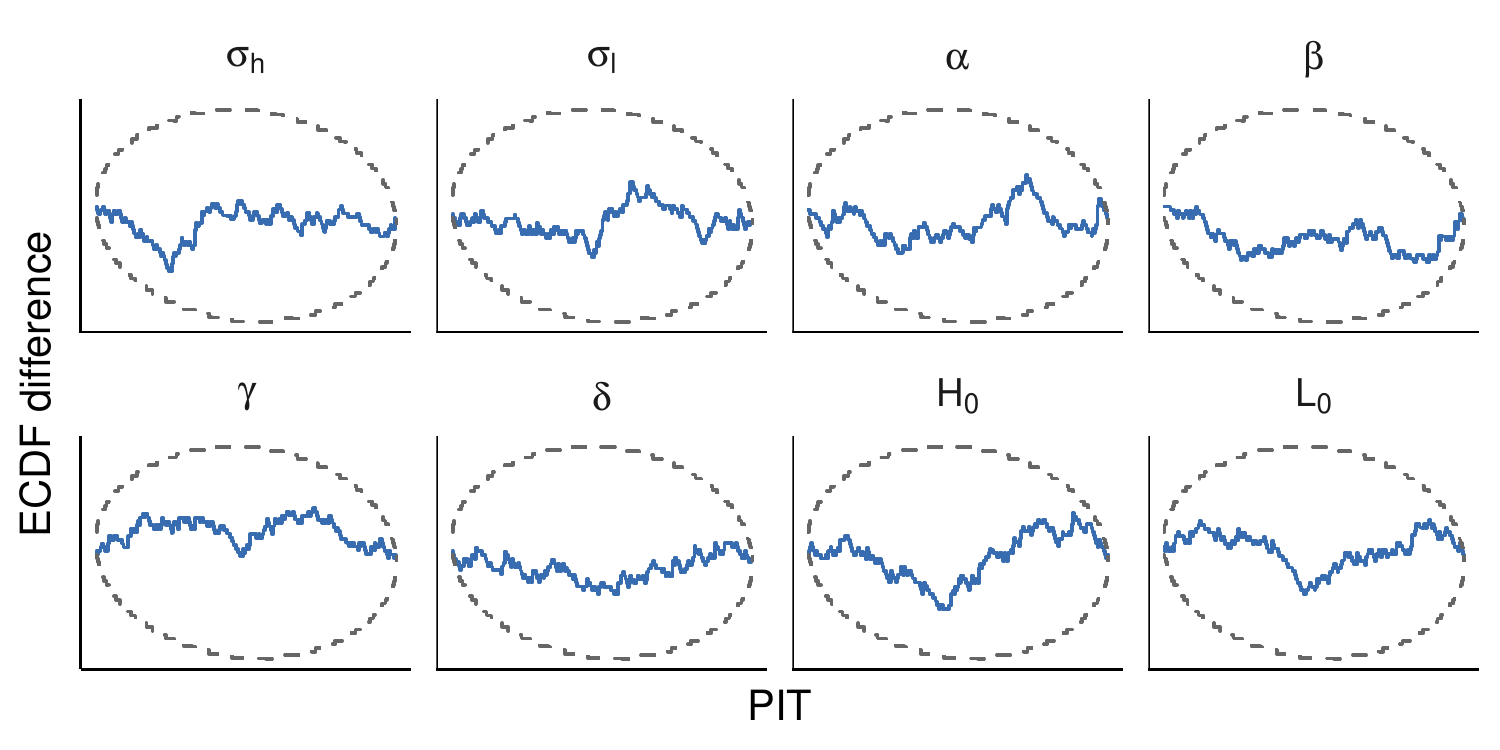}
    \caption{Individual parameter calibration assessments.}
    \label{fig:app:lv-prior-sbc-results}
    \end{subfigure}
    \begin{subfigure}[b]{0.9\linewidth}
    \centering
\includegraphics[width=1.0\linewidth]{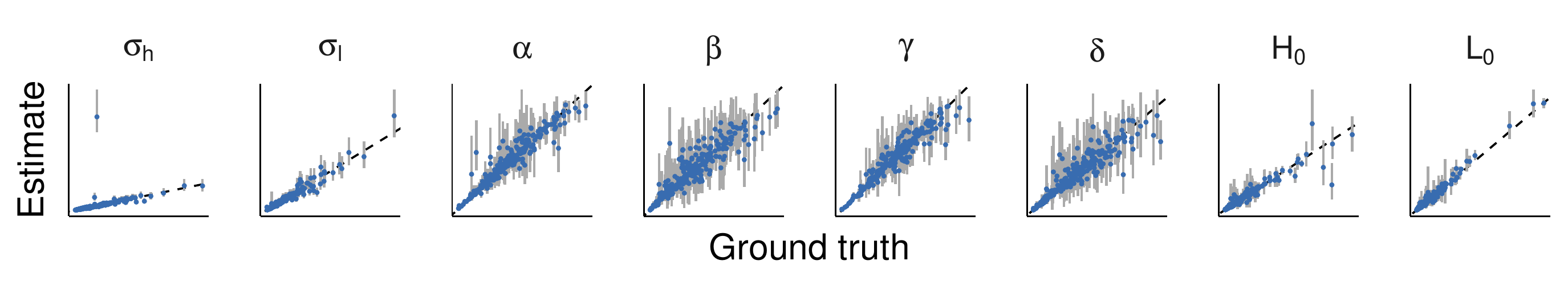}
    \caption{Parameter recovery.}
    \label{fig:app:lv-prior-sbc-recovery}
    
    \end{subfigure}
    \caption{Per parameter results of prior SBC for the Lotka-Volterra model.}
\end{figure}

In Figure~\ref{fig:app:lv-prior-sbc-results} we display the graphical calibration assessment of the individual parameters obtained with prior SBC for the Lotka-Volterra model. Even without correction for multiple comparison, the inference of the parameter posteriors looks well calibrated. In the parameter recovery plots for prior SBC, displayed in Figure~\ref{fig:app:lv-prior-sbc-recovery}, we see no systematic biases, but there are some cases where the posterior is very far from the ground truth.
As discussed in Section 3.2 of the article, these outliers are likely due to the posterior having distinct modes, which leads to a high probability of one or more MCMC chains not properly exploring the full posterior.

\begin{figure}[p]
    \centering
    \begin{subfigure}[b]{0.9\linewidth}
    \centering
    \includegraphics[width=.8\linewidth]{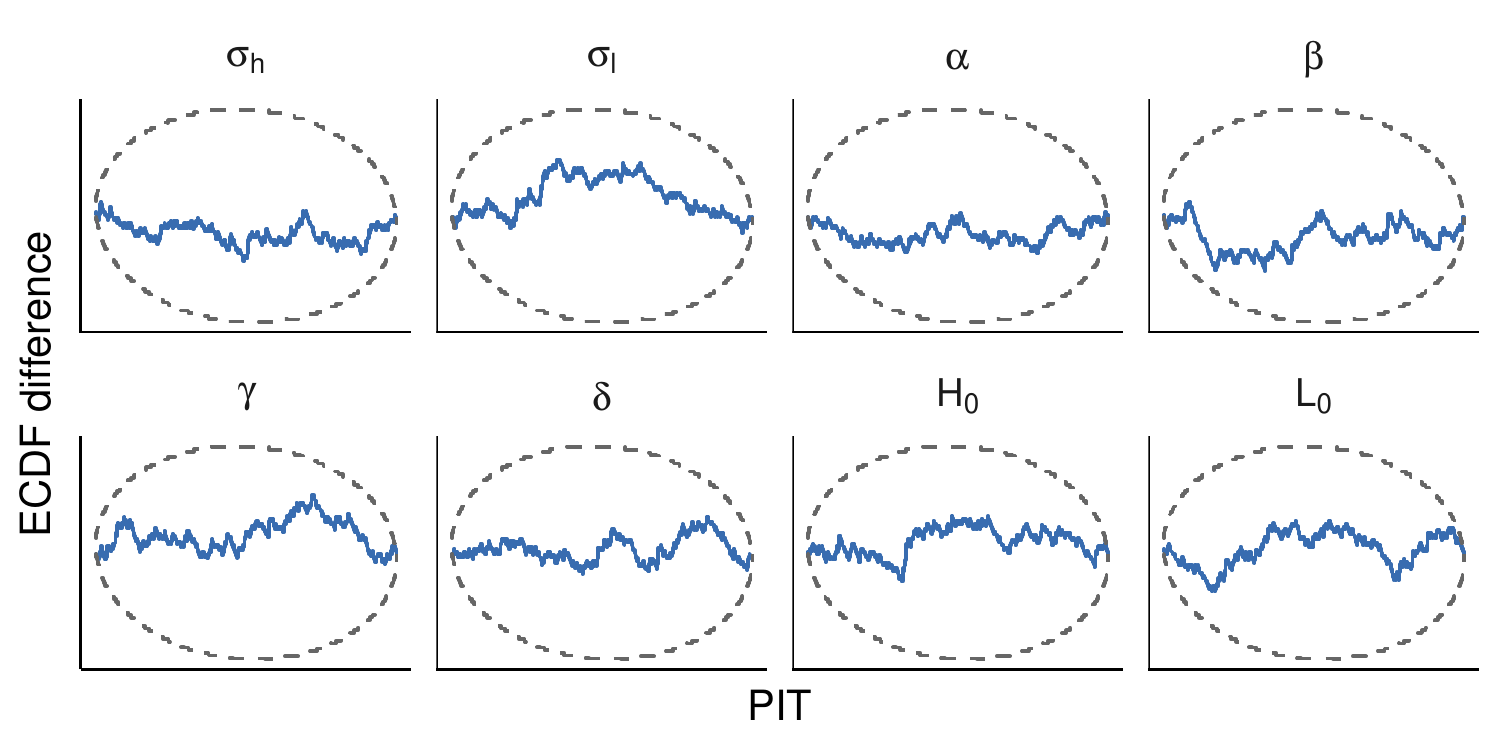}
    \caption{Individual parameter calibration assessments.}
    \label{fig:app:lv-posterior-sbc-results}
    \end{subfigure}
    \begin{subfigure}[b]{0.9\linewidth}
    \centering
    \includegraphics[width=1.0\linewidth]{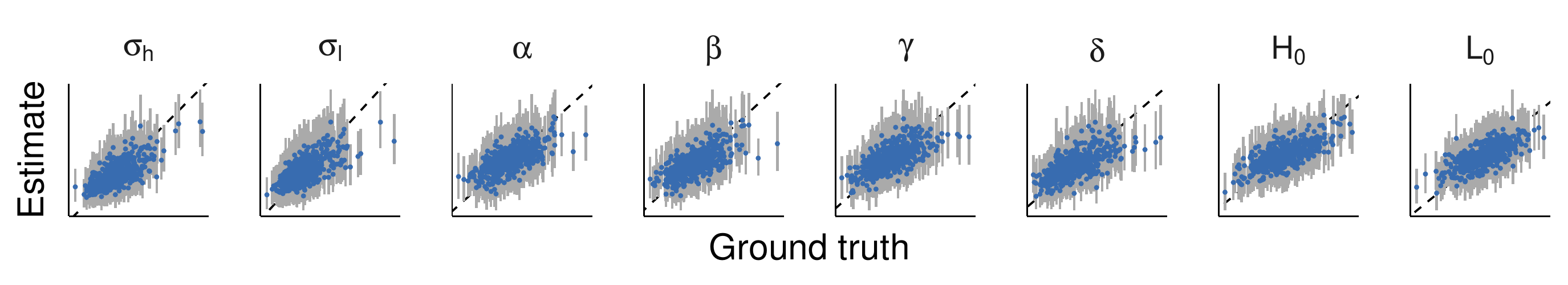}
    \caption{Parameter recovery.}
    \label{fig:app:lv-posterior-sbc-recovery}
\end{subfigure}
\caption{Per parameter results of posterior SBC for the Lotka-Volterra model.}
\end{figure}

Figure~\ref{fig:app:lv-posterior-sbc-results} and Figure~\ref{fig:app:lv-posterior-sbc-recovery} show the calibration assessment and recovery of the individual parameters once we condition the inference on the historical observations. Again, we observe no calibration issues. The parameter recovery also exhibits no outliers similar to those observed in prior SBC.

\FloatBarrier
\subsection*{Appendix B: Case study 3: Posterior SBC checking in amortized Bayesian inference}\label{app:abi}

In the following, we detail the full neural network settings as well as model specifications in case study 3, and present additional results concerning the parameter recovery of the amortized approximators.

\paragraph{Training setup and neural network settings}
The prior distribution $p(N)$ on the number of observations in the synthetic training datasets is an integer-valued uniform distribution $\mathcal{U}(50, 150)$.
The summary network is a DeepSet \citep{Zaheer2017} with mean-pooling, which combines equivariant and invariant neural network layers to extract a 32-dimensional summary vector $h_{\psi}(y)\in\mathbb{R}^{32}$ from each dataset $y\in\mathbb{R}^{N\times 3}$, where $N\in\{50,\ldots,150\}$ as specified above.
The inference network is an affine coupling flow with 6 coupling layers.
We train the neural network tandem for a total of 300 epochs with a batch size of 64, 1000 iterations per epochs, and determine the learning rate based on a cosine decay schedule with initial value of $5\cdot 10^{-4}$.

\paragraph{Model 1}
The first probabilistic model $M_1$ corresponds to ``model 2'' in \citet{ghaderi-kangavari_general_2023}.
It has the following data model,
\begin{equation}
    \begin{aligned}
        r_i, y_i & \sim \mathrm{DDM}(\alpha, \tau_{(e)i} + \tau_{(m)}, \delta, \beta),\\
        z_i & \sim \mathcal{N}(\gamma\cdot\tau_{(e)i}, \sigma^2), \\
        \tau_{(e)i} & \sim \mathcal{N}(\mu_{(e)}, s^2_{(\tau)}),
    \end{aligned}
\end{equation}
where $\mathrm{DDM}(\alpha, \tau, \delta, \beta)$ denotes a Wiener drift-diffusion model with threshold $\alpha$, non-decision time $\tau$, average drift rate $\delta$, and initial bias $\beta$. 
We use the following prior distributions from \citet{ghaderi-kangavari_general_2023}:
\begin{equation}
    \begin{aligned}
        \delta & \sim \mathcal{U}(-3, 3), \quad &
        \alpha & \sim \mathcal{U}(0.5, 2), \quad &
        \beta & \sim \mathcal{U}(0.1, 0.9), \quad \\
        \mu_{(e)} &\sim \mathcal{U}(0.05, 0.6), &
        \tau_{(m)}&\sim \mathcal{U}(0.06, 0.8), \quad &
        \sigma&\sim\mathcal{U}(0, 0.3),\quad \\
        s_{(\tau)}&\sim\mathcal{U}(0, 0.3), \quad &
        \gamma &\sim\mathcal{U}(0, 3).
    \end{aligned}
\end{equation}

\begin{figure}
    \centering
    \begin{subfigure}[b]{1.0\linewidth}
        \includegraphics[width=\linewidth]{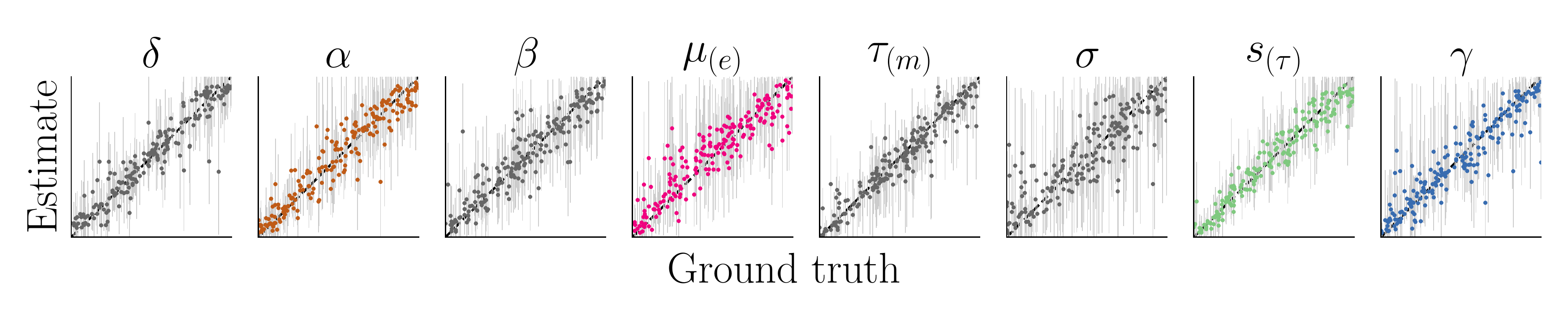}
        \caption{Recovery on synthetic datasets with $N=\Nobs$ observations each.}
        \label{fig:app:recovery-m1:N}
    \end{subfigure}
    \begin{subfigure}[b]{1.0\linewidth}
        \includegraphics[width=\linewidth]{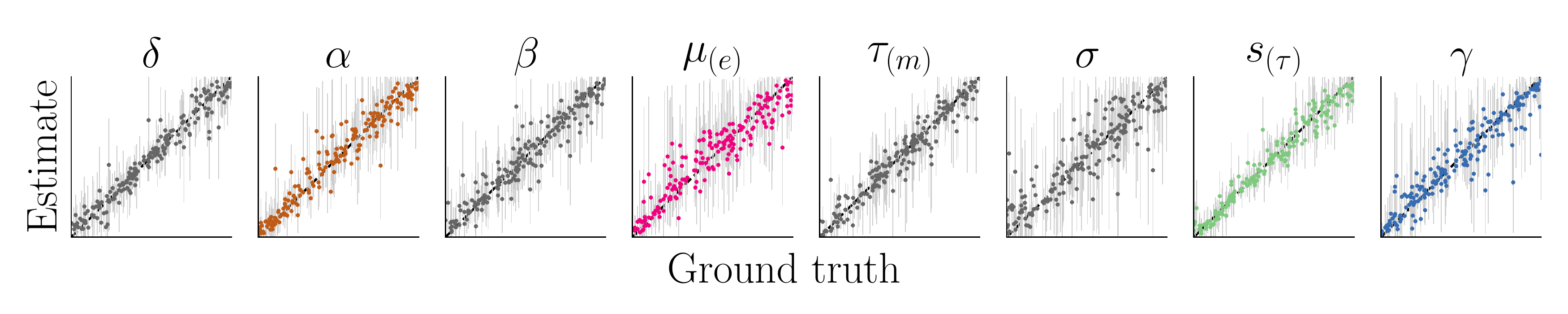}
        \caption{Recovery on synthetic datasets with $N=2\,\Nobs$ observations each.}
        \label{fig:app:recovery-m1:2N}
    \end{subfigure}
    \begin{subfigure}[b]{1.0\linewidth}
        \includegraphics[width=\linewidth]{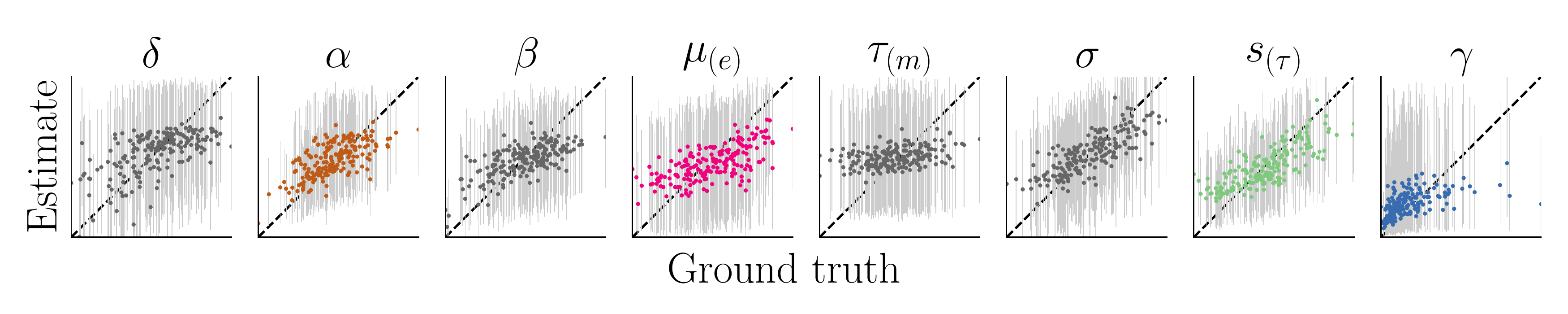}
        \caption{Recovery on data from the posterior predictive distribution conditional on $y^{(1)}_{\text{obs}}$.}
        \label{fig:app:recovery-m1:cond-yobs-1}
    \end{subfigure}
    \begin{subfigure}[b]{1.0\linewidth}
        \includegraphics[width=\linewidth]{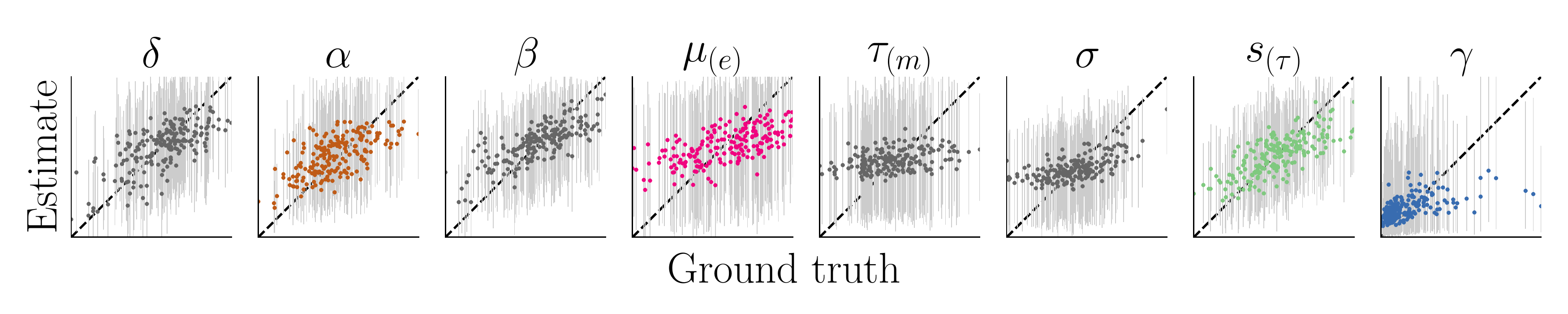}
        \caption{Recovery on data from the posterior predictive distribution conditional on $y^{(2)}_{\text{obs}}$.}
        \label{fig:app:recovery-m1:cond-yobs-2}
    \end{subfigure}
    \caption{Parameter recovery of the amortized approximator for model $M_1$.}
    \label{fig:app:recovery-m1}
\end{figure}

Figure~\ref{fig:app:recovery-m1} shows the parameter recovery of the amortized approximator across a range of settings.
We observe that most parameters can be recovered with sufficiently high accuracy given the number of available observations in each dataset (see Figure~\ref{fig:app:recovery-m1:N}).
Further, the epistemic uncertainty in the parameter estimates shrinks as the number of observations grows (see Figure~\ref{fig:app:recovery-m1:2N}).
Finally, errors in the posterior estimates propagate into the parameter recovery based on simulated data from the approximate posterior predictive distribution (see Figure~\ref{fig:app:recovery-m1:cond-yobs-1},\subref{fig:app:recovery-m1:cond-yobs-2}).
This evaluation follows the same principle as posterior SBC but reports the data-conditional parameter recovery rather than the data-conditional simulation-based calibration results.
In summary, the patterns from the empirical parameter recovery match the results from (prior and posterior based) simulation-based calibration checking in Section~2.3.

\begin{figure}[t]
    \centering
    \begin{subfigure}[b]{1.0\linewidth}
        \includegraphics[width=\linewidth]{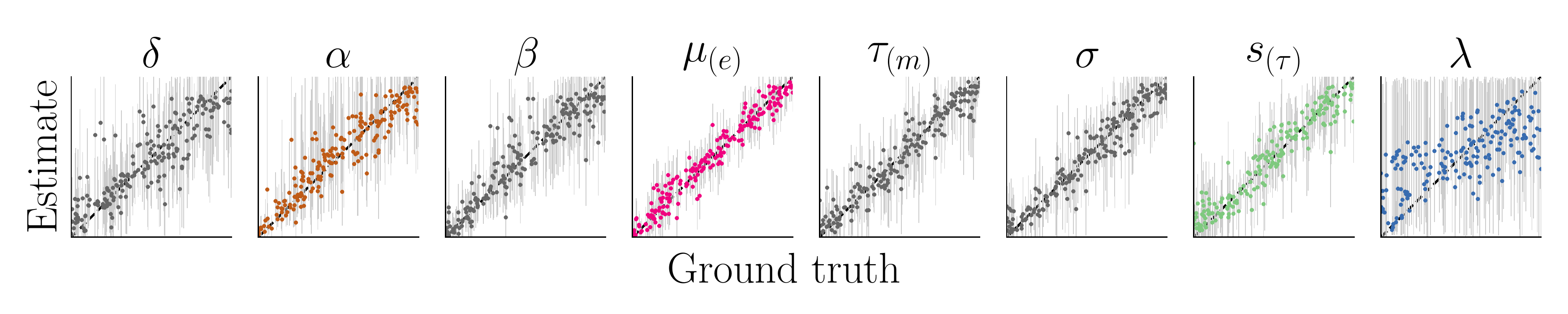}
        \caption{Recovery on synthetic datasets with $N=\Nobs$ observations each.}
        \label{fig:app:recovery-m2:N}
    \end{subfigure}
    \begin{subfigure}[b]{1.0\linewidth}
        \includegraphics[width=\linewidth]{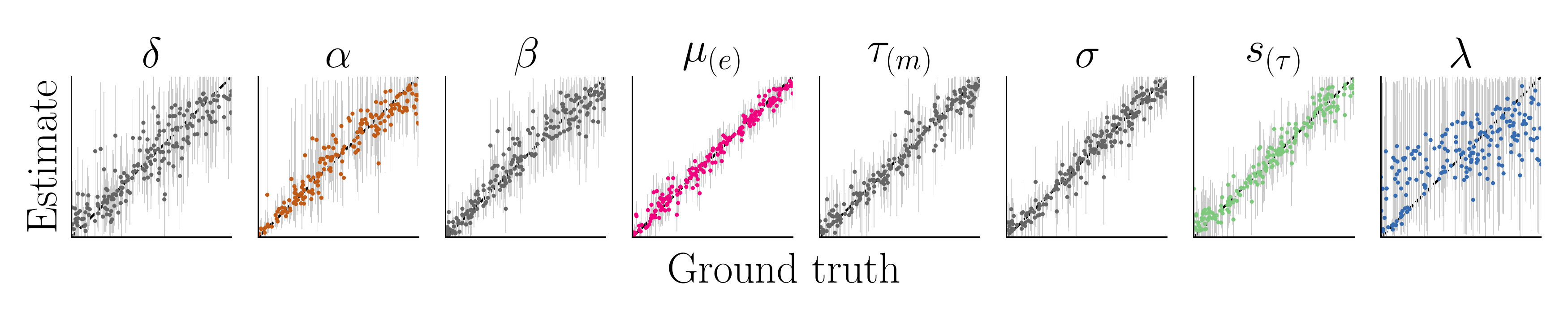}
        \caption{Recovery on synthetic datasets with $N=2\,\Nobs$ observations each.}
        \label{fig:app:recovery-m2:2N}
    \end{subfigure}
    \begin{subfigure}[b]{1.0\linewidth}
        \includegraphics[width=\linewidth]{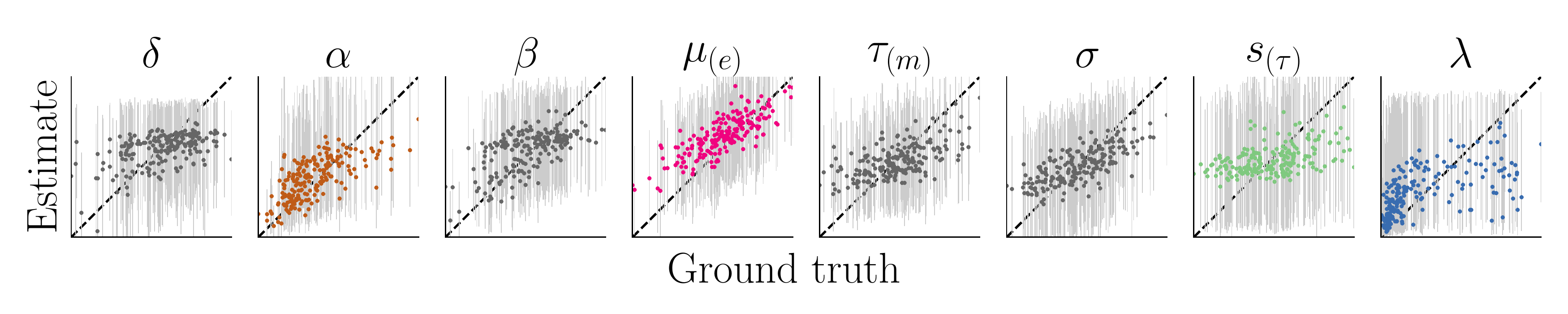}
        \caption{Recovery on data from the posterior predictive distribution conditional on $y^{(1)}_{\text{obs}}$.}
        \label{fig:app:recovery-m2:cond-yobs-1}
    \end{subfigure}
    \begin{subfigure}[b]{1.0\linewidth}
        \includegraphics[width=\linewidth]{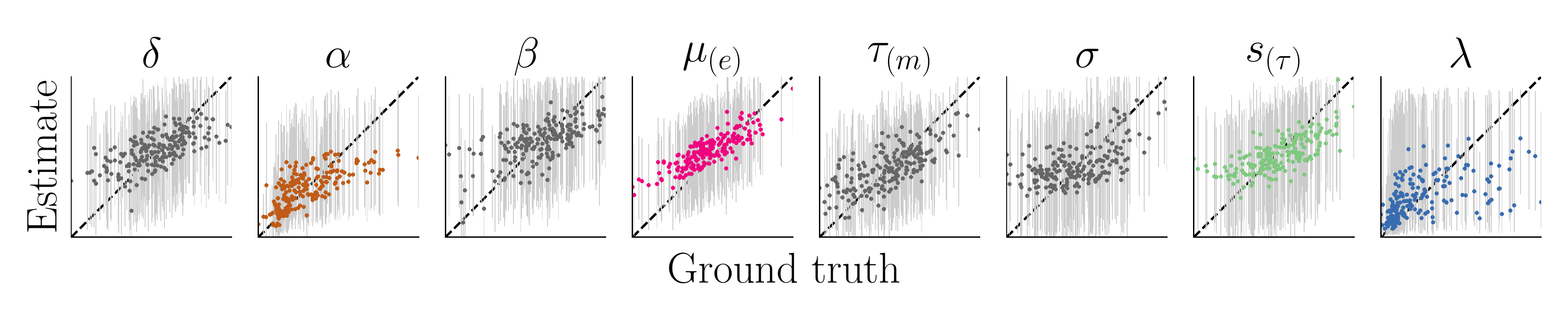}
        \caption{Recovery on data from the posterior predictive distribution conditional on $y^{(2)}_{\text{obs}}$.}
        \label{fig:app:recovery-m2:cond-yobs-2}
    \end{subfigure}
    \caption{Parameter recovery of the amortized approximator for model $M_2$.}
    \label{fig:app:recovery-m2}
\end{figure}

\FloatBarrier
\paragraph{Model 2} 
The second probabilistic model $M_2$ represents ``model 6'' by \citet{ghaderi-kangavari_general_2023}, which implements a drift-diffusion model with collapsing boundary \citep{Drugowitsch2012,Hawkins2015,Ratcliff2016}.
The collapsing boundaries are formalized through a scaled Weibull cumulative distribution function,
\begin{equation}
    \begin{aligned}
        u(t) & = \alpha - \left(1 - \exp\left(-\left(\dfrac{t}{\lambda}\right)^3\right)\right)\cdot(0.5\cdot \alpha),\\
        l(t)&=\alpha - u(t),
    \end{aligned}
\end{equation}
with upper threshold $u(t)$ and lower threshold $l(t)$ as a function of the time passed in the experiment.
Again, we follow \citet{ghaderi-kangavari_general_2023} and use the following prior distributions:
\begin{equation}
    \begin{aligned}
        \delta & \sim \mathcal{U}(-3, 3), \quad &
        \alpha & \sim \mathcal{U}(0.5, 2), \quad &
        \beta & \sim \mathcal{U}(0.1, 0.9), \quad \\
        \mu_{(e)} &\sim \mathcal{U}(0.05, 0.6), &
        \tau_{(m)}&\sim \mathcal{U}(0.06, 0.8), \quad &
        \sigma&\sim\mathcal{U}(0, 0.3),\quad \\
        s_{(\tau)}&\sim\mathcal{U}(0, 0.3), \quad &
        \lambda &\sim\mathcal{U}(0.5, 4).
    \end{aligned}
\end{equation}
Paralleling the previous evaluation for model $M_1$, Figure~\ref{fig:app:recovery-m2} reports the parameter recovery capabilities of the amortized approximator.
Again, most parameters can be recovered with sufficiently high accuracy given the number of available observations in each dataset (see Figure~\ref{fig:app:recovery-m2:N}), and the uncertainty shrinks when the datasets contain more observations (see Figure~\ref{fig:app:recovery-m2:2N}).
The large epistemic uncertainty in the estimate of the boundary collapse parameter $\lambda$ has previously been reported by \citet{ghaderi-kangavari_general_2023}.
Furthermore, the parameter recovery patterns mirror the results from (prior and posterior based) simulation-based calibration checking in Section~3.3.
Most notably, the previously reported data-conditional miscalibration in the parameter $\mu_{(e)}$ (Figure~10b in the article) manifests in a biased data-conditional parameter recovery in Figure~\ref{fig:app:recovery-m2:cond-yobs-1},\subref{fig:app:recovery-m2:cond-yobs-2}.

\end{document}